\documentclass[%
 aip, cha,
 amsmath,amssymb,
reprint,%
superscriptaddress,longbibliography,floatfix,showkeys
]{revtex4-2}

\usepackage{graphicx}% Include figure files
\usepackage{dcolumn}% Align table columns on decimal point
\usepackage{bm}% bold math
\usepackage[utf8]{inputenc}
\usepackage[T1]{fontenc}
\usepackage{mathptmx}
\usepackage{etoolbox}
\usepackage[normalem]{ulem}

\makeatletter
\def\@email#1#2{%
 \endgroup
 \patchcmd{\titleblock@produce}
  {\frontmatter@RRAPformat}
  {\frontmatter@RRAPformat{\produce@RRAP{*#1\href{mailto:#2}{#2}}}\frontmatter@RRAPformat}
  {}{}
}%
\makeatother

\usepackage[dvipsnames]{xcolor}
\usepackage{amsbsy}
\usepackage{tikz}
\usepackage{amsthm,mathrsfs,amsopn,amsfonts,amsmath,amssymb}

\usepackage[justification=RaggedRight]{caption}
\AtBeginEnvironment{matrix}{\everymath{\displaystyle}}
\DeclareMathOperator{\Var}{Var}

\usepackage{url}

\usepackage[colorlinks=true,linkcolor=blue,citecolor=blue,urlcolor=blue]{hyperref}

\usepackage[capitalise]{cleveref}
\usepackage{comment}

\newcommand{\md}[1]{\langle k^{(#1)} \rangle}
\newcommand{\MM}[1]{\textcolor{red}{[Martin: #1]}}
\newcommand{\ML}[1]{\textcolor{orange}{[Max: #1]}}

\newcommand{\rev}{\color{black}}

\begin{document}

\title{On the efficiency of pairwise Hamiltonian control to desynchronize the higher-order Kuramoto model}
%\title{Efficiency of pairwise Hamiltonian desynchronization control on the higher-order Kuramoto model}
%Higher-order interactions non-trivially affect the efficiency of the desynchronizing pairwise-based Hamiltonian control on the higher-order Kuramoto model
\author{Martin Moriamé}
\affiliation{Department of Mathematics and Namur Institute for Complex Systems (naXys), University of Namur, 5000 Namur, Belgium}
    \email{martin.moriame@unamur.be}
\author{Riccardo Muolo}
\affiliation{RIKEN Center for Interdisciplinary Theoretical and Mathematical Sciences (iTHEMS), Saitama 351-0198, Japan}
\affiliation{Department of Systems and Control Engineering, Institute of Science Tokyo (former Tokyo Tech), Tokyo 152-8552, Japan}
\author{Timoteo Carletti}
\affiliation{Department of Mathematics and Namur Institute for Complex Systems (naXys), University of Namur, 5000 Namur, Belgium}
\author{Maxime Lucas}
\affiliation{Department of Mathematics and Namur Institute for Complex Systems (naXys), University of Namur, 5000 Namur, Belgium}
\affiliation{Earth and Life Institute, Mycology, Catholic University of Louvain, B-1348 Louvain-la-Neuve, Belgium}

\date{\today}

\begin{abstract}
    Synchronization of coupled oscillators is observed in many natural and engineered systems and emerges due to the interactions within the system. It can be both beneficial, e.g., in power grids, and harmful, e.g., in epileptic seizures. In the latter case, efficient control methods to desynchronize the systems are crucial. Recent studies have shown that interactions are not always pairwise, but higher-order, i.e., many-body, and this greatly affects the dynamics. For instance, higher-order interactions increase the linear stability of synchronized states but simultaneously shrink their attraction basin, with potentially opposite effects on control methods. 
    Here, we use a minimally invasive pairwise control based on Hamiltonian control theory, and investigate its efficiency on phase oscillators with higher-order interactions.
    We show that, if the initial phases are close to the synchronized state, higher-order interactions make desynchronization more difficult to achieve. Otherwise, a non-monotonic effect appears: intermediate strengths of higher-order interactions impede desynchronization while larger ones facilitate it. In all cases, the control can desynchronize the system with a sufficient number of controlled nodes and intensity.
    %The synchronization of interacting units is crucial in many fields of nature and engineering, and its control and modulation are of huge importance. Even though it classically has to be strengthened and stabilized, some situations, e.g., Epilepsy in neuroscience, require its impediment. On the other hand, while the interactions between the units were classically modeled with pairwise networks, the recently renewed interest in higher-order interactions brought new perspectives as they dramatically affect the systems' synchronizability. In particular, they have been shown to increase the local stability of full synchronization and narrow its basin of attraction. In this work, we cross those recent discoveries with a minimally invasive Hamiltonian control method designed to desynchronize the pairwise Kuramoto model. We explore whether the addition of higher-order interactions is impeding or facilitating the desynchronization, i.e., which of the induced effects, deepening or shrinking of basin, is dominant. We show that, if the initial phases are close to the synchronized state, then the higher-order interactions make desynchronization more difficult to achieve. Otherwise, a non-monotonic effect appears: intermediate strengths of higher-order interactions impede desynchronization while larger ones facilitate it. In all cases, the control can desynchronize the system with a sufficient number of controlled nodes and control intensity.
\end{abstract}

\maketitle 
%-----------------------------------------------------------------------

\begin{quotation}
Many natural and engineered systems are composed of ensembles of interacting oscillators. When the interactions are strong enough, the oscillators may synchronize, meaning that they evolve with the same rhythm. While synchronization can be useful in many applications, there are important cases, e.g., in brain dynamics, in which it can be harmful and needs to be controlled and reduced. Pairwise interactions have been largely used to model coupled oscillators. However, increasing evidence shows that interactions are often higher-order, i.e., involving more than two oscillators at a time. Such interactions dramatically shape the dynamics, providing a plethora of synchronization patterns. In this work, we examine how such higher-order interactions influence the effectiveness of a simple control strategy designed to desynchronize a population of coupled oscillators. By studying different initial conditions and interaction strengths, we provide a clearer picture of when higher-order interactions make desynchronization harder or easier to achieve.
\end{quotation}

\section{Introduction}

%\the\linewidth$

Synchronization is one of the most astonishing examples of self-organization in complex systems. Since its description by Huygens in the 1600s, the synchronization of coupled oscillators has been observed in many natural and engineered systems, and finds applications from health and neuroscience to mechanics and power grids~\cite{Win80,pikovskysynchronization,strogatz2004,arenas2008synchronization,boccaletti2018synchronization}. 
Since Kuramoto proposed his canonical model with all-to-all coupling~\cite{kuramoto1975selfentrainment,kuramoto1984chemical}, the effect of complex network structures has been investigated in many extensions~\cite{strogatz2000kuramoto,acebron2005kuramoto}.
%Oscillatory systems are highly dimensional~\cite{Win80,strogatzbook} and, in general, systems of such coupled oscillators are not easy to analyze. However, the transition to synchronization can be studied by considering, instead, coupled phase oscillators, as correctly conjectured by Winfree in 1967~\cite{winfree1967biological}. Just a few years later, in 1975, Kuramoto was able to obtain a solvable model of coupled phase oscillators~\cite{kuramoto1975selfentrainment}, through which it was possible to fully describe the emergence of synchronization~\cite{kuramoto1984chemical,Win80}. Such model, nowadays called Kuramoto model, had been obtained by reducing the highly dimensional dynamics of each oscillator to a phase variable, via a technique called phase reduction~\cite{nakao16,pietras2019network,monga2019phase,kuramoto2019concept}. Note that the Kuramoto model makes sense as the phase description of an interacting system of limit cycles. In fact, the phase is not a physical variable \textit{per se}, but it is obtained via the phase reduction from the \textit{physical} state variables of the oscillatory system~\cite{kuramoto1984chemical,pikovskysynchronization}.
%
In general, synchronization can be beneficial to the system, for example, in cardiac dynamics\cite{Stefanovska2000} and power grids~\cite{totz2020control}; but it can also be harmful in other contexts, such as mechanics~\cite{dallard2001london}, aviation~\cite{livne2018aircraft,liu2018active,liu2025flutter}, or neuroscience~\cite{UHLHAAS2006neural,louzada2012suppress,wilson2014optimal,DBS}.  
For example, neurons are known to synchronize at the onset of epileptic seizures~\cite{Mormann2000MeanPhase,daSilva_et_al2003DynamicalDeseaseBrain}. In such cases, one wants to be able to desynchronize the system.
%Examples come from mechanics~\cite{dallard2001london}, aviation~\cite{livne2018aircraft,liu2018active,liu2025flutter}, or neuroscience~\cite{UHLHAAS2006neural,louzada2012suppress,wilson2014optimal,DBS}. {\color{green}[Ric: @Martin, I added only some basic references, can you add more?].}

Control theory offers several approaches to achieve synchronization including optimal control \cite{vinter2010optimal,monga2019optimal,fujii2025optimal}, pinning control~\cite{grigoriev1997pinning,sorrentino2007controllability,wang2002pinning,Wenwu2009onpinning,chen2014pinning,Turci2014adaptive,liu2021optimizing}, or adaptive control~\cite{Wang2008adaptive,Wenwu2009onpinning,Turci2014adaptive,liu2018active}, to name a few. Let us however observe that the literature dealing with desynchronization is more rare~\cite{gjata2017using,gambuzza2016pinning,asllani2018minimally,moriame2025onthelocation,muolo2025pinning}.
%Methods to achieve desynchronization are less common. 
%{[\color{red} Martin: I add some here and in the former paragraph. All concerning control of pairwise networks as the control papers for hypergraphs come after.]} From a control theory perspective, developing methods to achieve \emph{desynchronization} is rather unconventional and not many methods are readily available. 
One relevant research line aimed at desynchronizing coupled oscillators originated with Ref.~[\onlinecite{gjata2017using}]. It relies on the idea of embedding the Kuramoto model into a Hamiltonian system~\cite{witthaut2014kuramoto}, to then leverage Hamiltonian control theory~\cite{vittot2004perturbation,ciraolo2004control} to build a pinning control strategy aimed at desynchronizing the system. In Ref.~[\onlinecite{asllani2018minimally}], Asllani \textit{et al.} went a step further by introducing a modified control that results to be minimally invasive---the control magnitude is proportional to the level of synchronization and thus acts only when necessary---and requires less knowledge of the system parameters, making it easier to implement in practice.
%. Moreover, this second version of the control is potentially easier to implement in an applied case because it requires a less exhaustive knowledge of the system's parameters.

% removed because cited in first paragraph now
%Although early works on the Kuramoto model consider a pairwise and global (i.e., all-to-all) coupling between the phase oscillators~\cite{strogatz2000kuramoto,acebron2005kuramoto}, the interactions have then been modeled by more realistic complex networks~\cite{newmanbook,Latorabook}. The Kuramoto model in such a framework has been extensively studied and it was shown that the dynamics is much reacher and deeply affected by the complex topology~\cite{arenas2006synchronization,nicosia2013remote,rodrigues2016}. 

Recent evidence suggests that, in many systems, interactions tend to be higher-order, i.e., between more than two nodes, rather than just pairwise as in the traditional framework described above~\cite{battiston2020networks,battiston2021physics,bianconi2021higher,majhi2022dynamics,bick2023higher,boccaletti2023structure,muolo2024turing,millan2025topology}. Consequently, the interactions are more accurately described by generalization of pairwise networks, such as hypergraphs and simplicial complexes. Importantly, higher-order interactions have been shown to dramatically affect dynamics, e.g., with chaotic oscillators~\cite{Krawiecki2014,gambuzza2021stability,gallo2022synchronization}, swarming and active matter~\cite{anwar2024collective,anwar2025two,leon2025collective}, chimera states~\cite{kundu2022higherorder,muolo2024phase}, stochastic resonance~\cite{wang2025network}, pattern formation~\cite{carletti2020dynamical,muolo2023turing}, opinion dynamics~\cite{iacopini2019simplicial,neuhauser2020multibody,de2020social,deville2021consensus,alvarez2021evolutionary,lucas2023simplicially}, and random walks~\cite{schaub2020random,carletti2020random,febbe2026random}, to name a few. 

In systems of coupled phase oscillators, i.e., higher-order Kuramoto models~\cite{battiston2025collective}, 
%. Because such more complex interactions make the dynamics reacher, it was shown that, in general, synchronization is more difficult to achieve when the topology is higher-order
higher-order interactions have been shown to naturally arise from phase reduction~\cite{leon2019phase,gengel2020high,bick2024higher,mau2024phase,fujii2025emergence}, and to induce multi-stability, chaos, and explosive transitions, among others~\cite{tanaka2011multistable,bick2016chaos,skardal2019abrupt,skardal2020higher,millan2020explosive,lucas2020multiorder,kovalenko2021contrarians,adhikari2023synchronization,zhang2023higher,leon2024higher,fariello2024third,costa2024bifurcations,huh2024critical,wang2024coexistence,leon2025theory,namura2026optimal}. Recently, an interesting effect was described, that is of particular interest for this study: higher-order interactions can make the attraction landscape of synchronized states such as twisted states, ``deeper but smaller'', that is, increasing higher-order coupling strength increases linear stability but shrinks attraction basins~\cite{zhang2023deeper}. 

Early efforts have started to adapt control methods to modulate the synchronization of oscillators with higher-order interactions%
%A recent effort has been made by scholars to develop control methods to modulate synchronization in higher-order system
~\cite{chen2021controllability,de2022pinning,della2023emergence,de2023pinning,shi2023synchronization,rizzello2024pinning,xia2024pinning,li2024synchronization,muolo2025pinning,Moriame2025Hamiltonian}. In particular, building on Refs.~[\onlinecite{gjata2017using,asllani2018minimally}], a higher-order Hamiltonian control scheme was defined by embedding a higher-order Kuramoto model into a  Hamiltonian system~\cite{Moriame2025Hamiltonian}.
This higher-order version of the control was shown to be able to desynchronize the higher-order Kuramoto model better than the pairwise control, when higher-order interactions are not too weak. In cases where pairwise interactions are prominent, the higher-order control also worked, but the pairwise control was sufficient. However, the higher-order extension makes the control terms more expensive to compute and to implement; to obtain a simplified minimally invasive version of the latter seems to be a hard task to achieve. Thus, being able to control a higher-order system with the pairwise minimally invasive method developed in~[\onlinecite{asllani2018minimally}] remains of great interest. 

%Interestingly, the study~\cite{Moriame2025Hamiltonian} focused on situations in which the initial conditions were close to the synchronized state. In this context, the stability of the synchronized state is known to be reinforced by higher-order interactions, which justifies the need of the higher-order control extension.  However, it has also been proven in~\cite{zhang2023deeper} that the presence of higher-order interactions can reduce significantly the size of the basin of attraction.

In the context of the ``deeper but smaller'' effect induced by higher-order interactions, a natural question arises: how is the desynchronization control affected by higher-order interactions? 
Indeed, the increased linear stability would tend to make desynchronization harder, but the reduced size of the attraction basin would have the opposite effect. The result of these two competing effects on the efficiency of the control remains unclear. 

%To have a better idea of wether it is possible to desynchronize the hiher-order systems with a minimally invasive pairwise control method, one have to take into account this double effect the higher-order interactions have on the synchronizability of a system. This rise the more general question of how do higher-order interactions affect the efficiency of this desynchronizing control method. In other words, how much does the induced increasing of the local stability complexify the control's task, and how much does the induced basin of attraction shrinking help to achieve it? 

In the present work, we aim to answer this question. To do so, we assess the ability of a minimally invasive pairwise control~\cite{asllani2018minimally} to desynchronize systems composed of oscillators coupled with higher-order networks. We test the effect of both the intensity of higher-order interactions and the initial distance 
%between initial conditions and the fully synchronized state. 
to the center of the attraction basin. We first found that, for initial conditions starting close to the fully synchronized state, the stronger the higher-order interactions, the less efficient the control.
%, even though it still manages to decrease the system synchronization. 
On the other hand, for less synchronized initial conditions, we found a non-monotonic behavior. While intermediate magnitudes of higher-order interactions hinder the control efficiency, higher ones improve it. These results are consistent with the behavior of uncontrolled systems~\cite{zhang2023deeper,wang2025higher,muolo2025when,wang2025moderate,skardal2025mixed}.
%\MartinAdd{, where strong higher-order interactions enhance the local stability of synchronization but hinder the synchronization of trajectories arbitrary initial conditions.}

The manuscript is organized as follows. In \cref{sec:model} we present the higher-order Kuramoto model and the pairwise minimally invasive control method we adapted from~[\onlinecite{asllani2018minimally}]. In \cref{sec:Exp1,sec:Exp2} we develop and show the results of our numerical experiments where initial conditions are, resp., synchronized or randomly deviated from synchronization. We then wrap up and conclude in \cref{sec:conclusion}.

\section{The model} \label{sec:model}

Let us consider the following system of $N$ coupled phase oscillators~\cite{skardal2019abrupt,zhang2023deeper}
\begin{equation}\label{eq:HOKM}
    \begin{split}
        \dot{\theta_i} =& ~ \omega_i ~+~ \frac{K_1}{\md1}\sum_{j=1}^N A^{(1)}_{ij}\sin(\theta_j - \theta_i) \\
        +& ~ \frac{K_2}{\md2}\sum_{j,k=1}^N A^{(2)}_{ijk}\sin(\theta_j + \theta_k - 2\theta_i)~ +~\mu\delta_i p_i ,
    \end{split}
\end{equation}
where $\theta_i\in[0,2\pi]$ is the phase of the $i$-th oscillator, and $\omega_i$ its natural frequency, drawn from a normal distribution ${\bm{\omega}}\sim\mathcal{N}(0,\sigma^2)$ with variance $\sigma^2$.
Interactions are determined, at orders 1 and 2, by the respective adjacency tensors 
$\mathbf{A}^{(1)}$ and $\mathbf{A}^{(2)}$, and coupling strengths $K_1\geq 0$ and $K_2\geq 0$. 
The full coupling strengths are normalized by the associated 
mean degrees $\md1$ and $\md2$ so that $K_1$ and $K_2$ can be fairly compared.
%discarding the different number of terms entering into the summations.
%The two first terms of \eqref{eq:HOKM} represent the internal dynamics of a node and the $1^{st}-$order interactions with its neighbors, and the y constitute the classical Kuramoto model on a pairwise network. here, we consider non-identical oscillators, i.e., the natural frequencies are not equal but drawn from a distribution ${\bm{\omega}}\sim\mathcal{N}(0,\sigma^2)$. The third term represent the natural extension of the Kuramoto model to $2^{nd}-$order iteration, i.e., interactions that are defined through $2^{nd}-$order hyperedges (hree-bodies interactions). 
Finally, node $i$ is controlled by $p_i$ with strength $\mu\ge0$ if $\delta_i=1$ ($=0$ otherwise, namely the node is not directly controlled). The control term will be defined hereafter in \cref{eq:control}.
Note that here, for the sake of simplicity, we restrain higher-order interactions to three-body ones. %\MartinDel{Note also that, in general, phases are not modeling physical variable but are rather obtained via phase-reduction of another system. }

\Cref{eq:HOKM} without control (i.e., $\mu=0$) is an extension\footnote{Note that the three-body interaction can take two different forms~\cite{leon2025theory,battiston2025collective}: $\sin(\theta_j + \theta_k - 2\theta_i)$, i.e., $(1,1,-2)$ interaction~\cite{namura2026optimal}, and $\sin(2\theta_j - \theta_k - \theta_i)$, i.e., $(2,-1,-1)$ interaction~\cite{namura2026optimal}. Full synchronization in both higher-order Kuramoto models behaves in a qualitatively analogous way with respect to the variations of the coupling strengths~\cite{muolo2025when}. Here, for sake of simplicity, we only consider on the $(1,1,-2)$ interaction.} of the Kuramoto model~\cite{skardal2019abrupt,zhang2023deeper,leon2025theory}. 
Such models are known to synchronize under certain conditions,
% of the parameters, e.g., if $\mathrm{A}^{(1)}$ and $\mathrm{A}^{(2)}$ are dense enough then the phases $\theta_i(t)$ tend to adopt the same dynamics, and/or $K_1$ is larger than certain critical values; as discussed in the Introduction, the effects of $K_2$ on synchronization are not as straightforward as for~\cite{zhang2023deeper,muolo2025when} $K_1$. 
and the level of synchronization is typically characterized by the Kuramoto order parameter $R$, defined as
\begin{equation}\label{eq:OrderParam}
    Re^{\imath\Psi} = \frac{1}{N} \sum_{j=1}^N e^{\imath\theta_j}\, ,
\end{equation}
where $\imath=\sqrt{-1}$ is the imaginary unit. The value of $R$ lies in $[0,1]$, with $R=0$ corresponding to a uniform distribution of the phases in $[0,2\pi]$ and $R=1$ to perfect synchronization in which all phases $\theta_i$ are equal. 
Note that the latter is an equilibrium only for systems with identical frequencies $\omega_i=\omega$. For distributed frequencies, a state in which $R$ is close to 1 is stable instead; we refer to it as the fully synchronized state. 
%The larger $K_1$, or the narrower the natural frequencies distribution, the larger is the maximal value of $R$.
Note that the Kuramoto model describes the dynamics of phase oscillators, which may be restrictive for potential applications, given that the phase is not a physical variable, i.e., it cannot be directly measured but needs to be extracted. However, this is what makes the Kuramoto model general: in fact, it can be obtained via the phase reduction of coupled limit cycle oscillators~\cite{kuramoto1984chemical,nakao16,pietras2019network,monga2019phase,kuramoto2019concept}, making it a good approximation of many oscillatory systems.

Finally, the last term denotes the feedback pinning control, where $p_i$ is defined as 
\begin{equation}\label{eq:control}
    p_i = -\frac12 K_1^2\frac{M^2}{\md1^2}R_M\tilde{R}_{M,i}\cos(\Psi_M-\tilde{\Psi}_{M,i})\, ,
\end{equation}
where $M=\sum_{i=1}^{N}\delta_i$ is the number of controlled nodes,
and where we defined the order parameter \eqref{eq:OrderParam} restricted to the $M$ controlled nodes,
 %The factor $\delta_i$ is equal to 0 or 1, depending whether the node $i$ is controlled or not, $\mu$ is a positive parameter characterizing the control intensity, and the third factor is defined as
%
\begin{equation}\label{eq:RM}
    R_Me^{\imath\Psi_M} = \frac{1}{M} \sum_{j=1}^N \delta_je^{\imath\theta_j}\, ,
\end{equation}
and a slight variation normalized by frequency differences
\begin{equation}\label{eq:RM_tilde}
    \tilde{R}_{M,i}e^{\imath\tilde{\Psi}_{M,i}} = \frac{1}{M} \mathop{\sum_{j=1}^N}_{\rev j\neq i} \delta_j\frac{e^{\imath\theta_j}}{\omega_j-\omega_i}\, .
\end{equation}
%
%The control measures the phases of this subset of the nodes, computes the control action from them, and re-injects it into their dynamics.
%
The control term~\eqref{eq:control} is a slightly modified version of the minimally invasive term proposed by Asllani \textit{et al.}~\cite{asllani2018minimally} so that it works on the same principle but is a little easier to use in practical cases (see~\cref{app:control} for details). {\rev In short, the control measures the phases of the $M$ pinned nodes, computes the control action which is proportional to their mean phase shifted by $\frac 14$ of its period, and re-injects it into their dynamics. Its magnitude is proportional to $R_M$ and $K_1^2$, so that it is large when the system needs to be desynchronized and negligible otherwise.} Model~\eqref{eq:HOKM} thus describes the interplay between two antagonistic forces: interactions tend to stabilize full synchronization (at least locally), provided that $K_1$ and $K_2$ are large enough, but the pinning control term tends to destabilize it. {\rev Let us note that this control method differs from classical pinning control, where the action on node $i$ depends only on its own dynamics. For instance, in control methods aiming to synchronize the system, the control action is often proportional to the distance between $\theta_i$ and a target trajectory to which the whole system has to converge~\cite{liu2021optimizing}. Here, the control acts to get $R_M\approx0$ so that total desynchronization can eventually be achieved. Hence, $p_i$ is necessarily a function of all the $M$ phases.}

Because we use the minimally invasive pairwise Hamiltonian control, we overcome a first issue related to the computational cost required to obtain a many-body control term as in~Ref. [\onlinecite{Moriame2025Hamiltonian}]. On the other hand, as we do not take into account higher-order terms in the control, the latter is not guaranteed to be efficient if $K_2$ is large in comparison with $K_1$~\cite{Moriame2025Hamiltonian}. In the following, we will benchmark our working assumption and determine in which cases the minimally invasive pairwise method proves to be sufficient to remove enough synchronization.

To measure the efficiency of the control, we use the time averaged order parameter
\begin{equation}\label{eq:AsymOrederPamram}
    \hat R := \langle R(t) \rangle_{t \in [t_i,t_f]}\, ,
\end{equation}
where the average has been computed by evaluating $R(t)$ in the interval $[t_i,t_f]$, with $t_i$ large enough to remove the initial transient dynamics.
The lower the value of $\hat R$, the better will be the control at reducing the synchronization, hence to result more efficient.

In addition, one can be interested in measuring frequency synchronization, i.e., phase locking. Indeed, in cluster states or twisted states, the phases are desynchronized, $R(t)\approx0$, but the effective frequencies are (almost) equal. To distinguish these states from complete incoherence, we use the following frequency order parameter~\cite{newman2025order}
\begin{equation}
    R_{\dot{\theta}}(t) = \max\left( 1 - \sqrt\frac{\Var(\dot{\bm{\theta}}(t))}{\Var({\bm{\omega}})} , 0\right)\, ,
\end{equation}
that lies in $[0,1]$. This quantity compares the variance of the effective frequencies to that of the natural frequencies. Similarly to the Kuramoto order parameter \cref{eq:OrderParam}, $R_{\dot{\theta}}(t)$ tends to 1 if the frequencies are synchronized---i.e., the effective frequencies are much less spread than the natural frequencies $\bm{\omega}$---but tends to 0 if they are not---i.e., they are at least as spread as $\bm{\omega}$. 
Similarly to phase synchronization, we report the averaged frequency order parameter
\begin{equation}
    \hat R_{\dot{\theta}} := \langle R_{\dot{\theta}}(t) \rangle_{t \in [t_i,t_f]}\, .
\end{equation}
As we will see below, the two order parameters will be complementary because the control can achieve full incoherence, $R(t)\approx0$ and $\hat R_{\dot{\theta}}\approx0$, but it can also desynchronize the phases without desynchronizing the frequencies, $R(t)\approx0$ but $\hat R_{\dot{\theta}}>0$.
This is because the control term $p_i$ vanishes as $R(t)\approx0$.
%even though the control does not aim to break frequency synchronization - e.g., if the controlled nodes are in a twisted state, then $R_M = 0$ and $p_i = 0$ - it can still be considered as a kind of order. It is, therefore, interesting to measure frequency synchronization to examine if such states can emerge after the synchronization breaking. 
%

%$R_{\dot{\theta}}(t)=1$ means that frequencies are equal and perfectly synchronized at time $t$, and $R_{\dot{\theta}}(t)=0$ that the frequencies' variance is greater or equal to the variance of the natural frequencies' distribution. In other words, in this case the frequencies are at least as desynchronized as in an uncoupled system, which means that they can fairly be considered as ``completely desynchronized''.

% \begin{comment}
%     \textcolor{red}{TO REMOVE?} On the other hand, it can be defined as the required time so that $R(t)$ is lower than a certain threshold $\tilde R$,. We thus introduce
% \begin{equation}\label{eq:ThresholdTime}
%     t_{\tilde R} := \min_t\{t: R(t)\leq \tilde R\}\, ,
% \end{equation}
% where $\tilde R \in [0,1]$ is fixed to an acceptably low value. The lower $t_{\tilde R}$, the more efficient is the control.
% \end{comment}

\section{The case of synchronized initial conditions}\label{sec:Exp1}

We now examine how the presence of higher-order interactions affects the efficiency of the control. 
%One way to assess this question is to investigate if the increasing intensity of higher-order interaction induces an increasing or decreasing of the required control strength to desynchronize the system. The aforementioned intensity of the higher-order interactions can be measured by $\md2$ and $K_2$. Here, the former is fixed, so this feature will only be measured by the latter. In this Section, we also fix $M=N$, i.e., we consider that all nodes are pinned, and the only parameter controlling the control intensity is $\mu$.
To do this, we measure how $\hat{R}$ and $\hat{R}_{\dot\theta}$ change as we vary the triadic coupling strength $K_2$ and control strength $\mu$. For each value of $K_2$ and $\mu$, we initialize the system by starting from perfectly synchronized phases $\theta_p = (0,\dots,0)^T$, with $R(0)=1$, and let the uncontrolled system, $\mu=0$, converge to the fully synchronized state $\theta_{\rm fs}$ after a transient time {\rev $t_{tr}$, with $0\ll$ }$R(t_{tr})<1$ because frequencies are not identical {\rev with} each other. Then, we activate the control, $\mu>0$, and let the system to converge to its stationary state. Finally, we measure $\hat{R}$ and $\hat{R}_{\dot\theta}$ for that final state, and use them as proxy of control efficiency. In the following, we vary $K_2$ from $0$ to $20$ and $\mu$ from $0$ to {\rev $1/8$. For simplicity,} we start by controlling all nodes, $\delta_i=1$ for all $i=1,\dots,N$.
The other parameters are set to $N=100$, {\rev $t_{tr}=250$, $t_i = t_{tr} + 190$, $t_f = t_{tr} + 200$,} $K_1=1$ and $\sigma=0.1$. The chosen hypergraphs topologies are the ones of the hyperrings (see \cref{ssec:Exp1Ring}) and random Erd\H{o}s-Rényi hypergraphs (see \cref{ssec:Exp1Random}).

%We consider fully synchronized initial conditions. It has to be noted that, in this work, "full synchronization" does not mean that all phases are exactly equal, i;e., perfect synchronization. The later is not a steady state of the \eqref{eq:HOKM}, because the oscillators have non-identical natural frequencies. Here, the fully synchronized state is rather a phase-locking state where all phases are very close to each other, and then $R$ is very close but not exactly equal 1. The exact shape of this phase-locking state depends on $\bm{\omega}$, $K_1$ and $K_2$. The larger $K_1$ and $K_2$ and the narrower $\bm{\omega}$, the closer is $R$ to 1. Again, let us stress that $K_2$ has such an effect because the system has initial conditions in the synchronized state. To take that into account, we start with preliminary conditions $\theta_p = (0,\dots,0)^T$ (i.e., a perfectly synchronized state) and let the uncontrolled system converge to the fully synchronized state $\theta_{\rm fs}$. Then, we set $\theta_0:=\theta_{\rm fs}$ and activate the control.

We observe here-after that the larger $K_2$, the larger the required $\mu$ value that guarantees  desynchronization. This means that, when starting from the fully synchronized state, at the center of its attraction basin~\cite{zhang2023deeper}, the increasing of the local stability is the dominant effect and it induces the loss of control's efficiency. In all cases, however, the pairwise minimally invasive control can desynchronize the system{\rev, both in phases and frequencies,} if $\mu$ is large enough.

\begin{figure*}[thb]
    \centering
    \includegraphics[width=\textwidth]{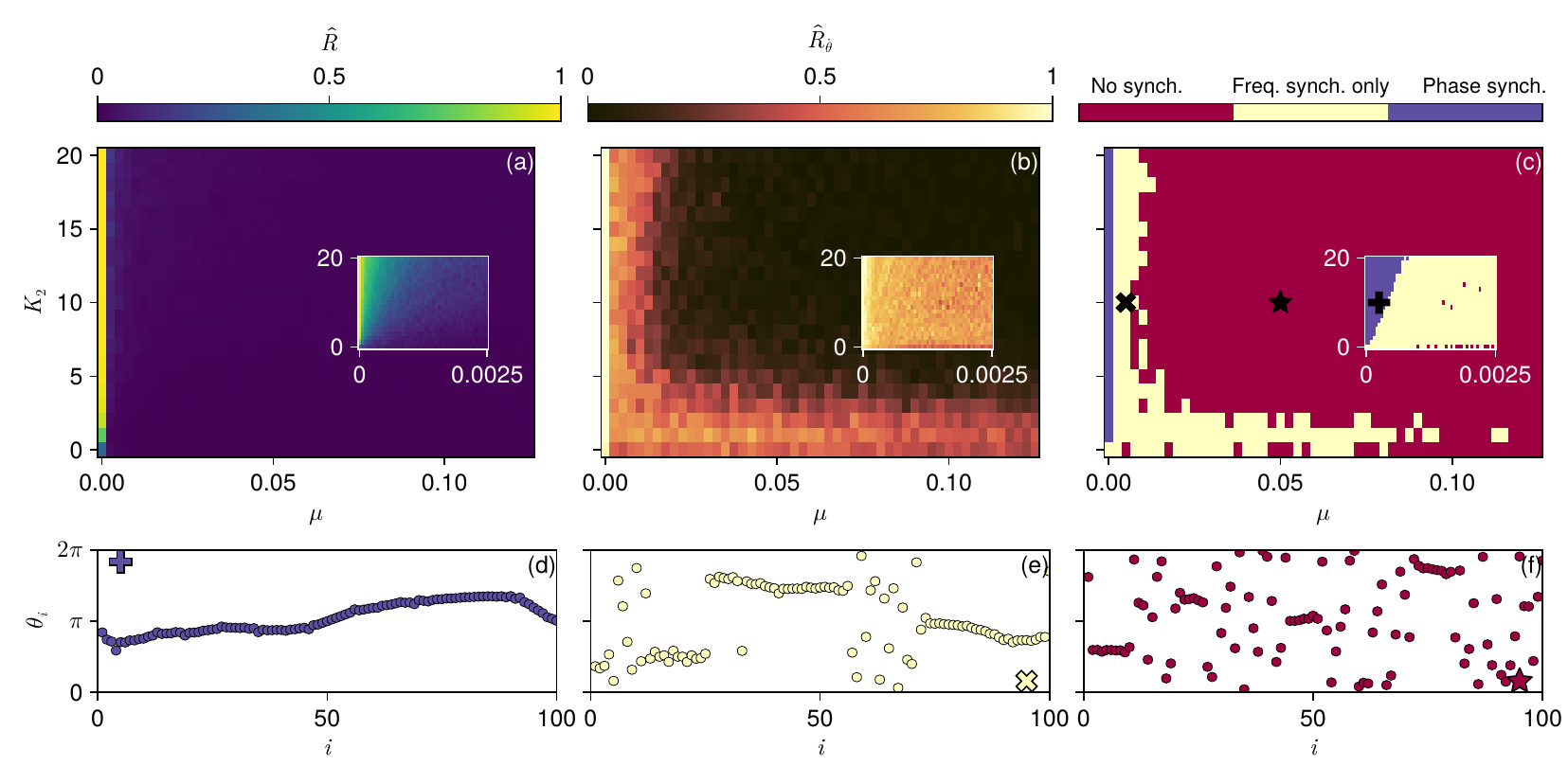}
    \caption{\textbf{Desynchronizing phase oscillators on a hyperring.}
    We show (a) the level of phase synchronization $\hat R$, and (b) of frequency synchronization $\hat{R}_{\dot\theta}$ as functions of triadic coupling strength $K_2$ and control strength $\mu$. 
    For each value of ($\mu$,$K_2$), we show the value averaged over 50 frequency distributions $\bm{\omega} \sim \mathcal{N}(0,0.01)$. 
    Insets zoom in on the low $\mu$ region.
    Panel (c) is obtained by merging the information from panels (a) and (b) and shows the three regimes of the controlled system---phase synchronization (blue, $\hat{R}\geq 0.4$), only frequency synchronization (yellow, $\hat{R}<0.4$ and $\hat{R}_{\dot\theta}\geq 0.5$), and no synchronization (red, otherwise)---and associated to the couples $(\mu,K_2)$. Panels (e)-(g) show example snapshots of these regimes as identified with the symbols: plus, cross and star.
    Parameters are set to $N=100$, $K_1=1$ and $r=2$.
    }
    \label{fig:K2_v_mu}
\end{figure*}

\subsection{Hyperrings}\label{ssec:Exp1Ring}

First, we focus on hyperring structures. Given $N$ nodes and a range $r$, a pairwise edge is drawn between node $i$ and each of its $2r$ nearest neighbors on the ring ($r$ on each side). A 2-hyperegde is also drawn between node $i$ and any two distinct nodes among these same neighbors. On this structure, system~\eqref{eq:HOKM} can be written 
\begin{equation}\displaystyle\label{eq:HOKMRing}
    \begin{split}
        \dot{\theta_i}&= \omega_i + \frac{K_1}{2r} \sum_{j=i-r}^{i+r} \sin(\theta_j-\theta_i) \\
        &+ \frac{K_2}{2r(r-1)} \sum_{j=i-r}^{i+r}\mathop{\sum_{k=i-r}^{i+r}}_{\rev k\neq j} \sin(\theta_j+\theta_k-2\theta_i) + \mu \delta_i p_i\, ,
    \end{split}
\end{equation}
where we used $\md1=2r$ and $\md2=2r(r-1)$, and the control term $p_i$ is still defined by \eqref{eq:control} and we set $\delta_i=1$ for all $i$. %Note that this hyperring structure is not a simplicial complex.

Due to the rotational invariance of the underlying substrate, the hyperring supports twisted states~\cite{wiley2006size,zhang2023deeper}. In fact, for pure pairwise interactions $K_2=0$, we know that full synchronization and twisted states are the only fixed points of the uncontrolled system $\mu=0$~\cite{wiley2006size}. For the uncontrolled system $\mu=0$, Ref.~[\onlinecite{zhang2023deeper}] also showed that, by increasing $K_2>0$, three things occur: the linear stability of twisted states increases, their basin of attraction shrinks, and new disordered states appear and become stable. 

We now consider the controlled system, $\mu>0$. In \cref{fig:K2_v_mu}, we report $\hat{R}$ and $\hat{R}_{\dot\theta}$ as functions of  the triadic coupling strength $K_2$ and the control strength $\mu$, for the case of $r=2$. We observe that phase synchronization $\hat{R}$ decreases as $\mu$ is increased, but increases as $K_2$ is increased (\cref{fig:K2_v_mu}a). The effect associated to $\mu$ is expected---stronger control desynchronizes the {\rev phases} better.
%---and the effect of $K_2$ is consistent with the literature 
The effect of $\mu$ and $K_2$ is similar on frequency synchronization: its increase tends to decrease $\hat{R}_{\dot\theta}$ (\cref{fig:K2_v_mu}b). 
There is nonetheless a difference: $\hat{R}_{\dot\theta}$ remains high for values of $K_2$ up to 5.

This phenomenology can be classified into three regimes of increasing disorder, as shown in \cref{fig:K2_v_mu}c. First, for small $\mu$ values, i.e., weak control, and regardless of $K_2$, the system remains in its initial synchronized regime (blue)---both, phases and frequencies are synchronized---as indicated by large $\hat{R}$ and large  $\hat{R}_{\dot\theta}$. In this phase synchronization regime, phases cluster around a single value and evolve with a common frequency (\cref{fig:K2_v_mu}d). Second, for intermediate values of $\mu$, phases are desynchronized but frequencies are still synchronized (yellow region), as indicated by low $\hat{R}$ but large  $\hat{R}_{\dot\theta}$. This configuration of the two order parameters can represent regimes such as twisted states and cluster states, in which phases are different but oscillators evolve at the same speed. In this frequency synchronization regime, phases appear to cluster around multiple values, but still evolve with a common frequency (\cref{fig:K2_v_mu}e). Notice that, even though this case is close to a cluster state, it is only partially clustered: a proportion of the oscillators display incoherent phases. Generalized order parameters further measure how clustered these states are (see \cref{fig:K2_v_mu_R2_R3} in Appendix~\ref{app:SuppFig}). Third, for strong control, that is, large $\mu$, no synchronization remains (red region), as indicated by low $\hat{R}$ and low $\hat{R}_{\dot\theta}$. In this disordered regime, phases evolve at different frequencies and as a result, phases are not coherent (\cref{fig:K2_v_mu}f). In \cref{fig:K2_v_mu}, we classify a given state into the three regimes as follows: if $\hat{R} \ge R_{\rm thr.}$, then we are dealing with phase synchronization; if $\hat{R} < R_{\rm thr.}$ but $\hat{R}_{\dot\theta} \ge 0.5$, then we are dealing with frequency synchronization only; lastly, if both $\hat{R} < R_{\rm thr.}$ and $\hat{R}_{\dot\theta} < 0.5$, we consider the system not to be synchronized, neither with respect to phase nor to frequency. We set $R_{\rm thr.}=0.4$ based on a preliminary visual inspection of the available data, so that the classification matches actual regimes. 
The results are qualitatively equivalent in the case $r=3$ (\cref{fig:k2_v_mu_r=3} Appendix~\ref{app:SuppFig}).
Overall, increasing the control strength $\mu$ increases desynchronization, as expected. 

\begin{figure}[t]
    \centering
    \includegraphics[width=\linewidth]{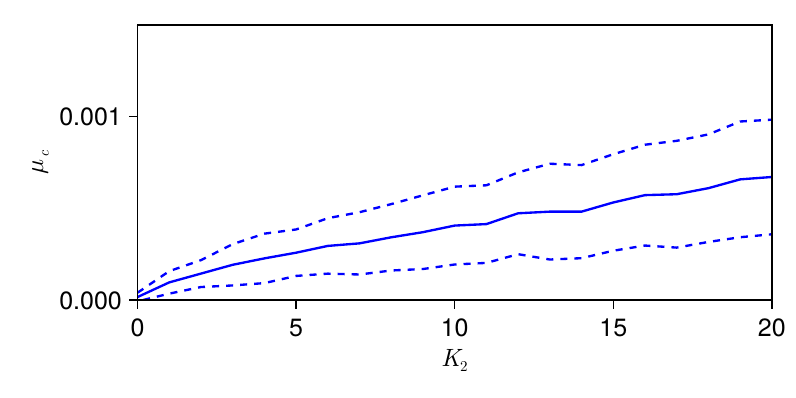}
    \caption{
    \textbf{Minimum control strength $\mu_c$ required to desynchronize $\hat R$ < 0.4 increases with triadic coupling strength $K_2$ on hyperring.} 
    We show $\mu_c$ averaged (solid line) over 50 frequency distributions $\bm{\omega}$, and one standard deviation away (dashed lines).
    Parameters are set to $N=100$, $K_1=1$ and $r=2$, as in \cref{fig:K2_v_mu}.}
    \label{fig:mu_c}
\end{figure}

To further characterize the effect of higher-order interactions on control efficiency, we define a critical control strength $\mu_c$ as the smallest $\mu$ such that $\hat R < R_{\rm thr.}$, that is, such that phases are sufficiently desynchronized. \Cref{fig:mu_c} shows that  $\mu_c$ grows with $K_2$, indicating that stronger control is needed to achieve a given level of phase desynchronization as $K_2$ increases, i.e., as triadic interactions are stronger. Although we know that by increasing $K_2$ tends to increase the linear stability of synchronization but shrinks its basin size, this monotonic trend suggests that the former effect is dominant over the latter if the initial conditions are synchronized. {\rev In \cref{app:SuppFig} (see \cref{fig:K2_v_mu_varying_R_thr}), we display analogous results obtained by varying the value of $R_{\rm thr}$, which leads to equivalent results.}
%as well as in \cref{fig:K2_v_mu}(c).

%On panels (b) and (c) one can observe that, although $\hat R$ rapidly decreases to 0 with $\mu$ increasing, $\hat{R}_{\dot\theta}$ does so more slowly. Although it decreases with $\mu$ and is close to 0 for large enough $\mu$, the control strength required to achieve frequency desynchronization is larger than for phase desynchronization. 

% \ML{include somewhere} It does not exactly look like a twisted state, but more like a kind of clustered state. However,  indicates that there is no notable presence of clustered-synchrony states in the trajectories. 

Additionally, we observe that the control strength $\mu$ necessary to achieve a given level of frequency desynchronization is larger for weaker triadic interactions, $K_2 \le 5$. This can be seen with $\hat{R}_{\dot\theta}$ (\cref{fig:K2_v_mu}b) and with the yellow region extending to larger $\mu$ (\cref{fig:K2_v_mu}c). 
This behavior may be explained by the fact that, in the uncontrolled system, twisted states are known to have large basins of attraction for small-to-moderate $K_2$, which shrinks at larger $K_2$~\cite{zhang2023deeper}.
As previously mentioned, note that the minimal control \eqref{eq:control}, by design, does not directly suppress frequency synchronization. Indeed, the control vanishes if the phases are fully desynchronized, $R=0$. Therefore, twisted sates and clustered states, which are characterized by $R\approx0$ and $R_{\dot\theta}\approx1$, are steady states of the controlled system.

In summary, the pairwise Hamiltonian control considered is able to suppress both phase and frequency synchronization, with phase desynchronization being achieved more readily due to the nature of the control scheme. For synchronized initial conditions, we find that by increasing the higher-order coupling strength $K_2$ systematically hinders phase desynchronization, as quantified by the monotonic increase of the critical control amplitude $\mu_c$. Frequency synchronization is even more robust for moderate $K_2$, owing to hypergraph symmetries and to the fact that it is not directly targeted by the control. Interpreting $\mu_c$ as an escape threshold from the synchronized basin, these results indicate that the enhancement of local linear stability induced by higher-order interactions dominates over the simultaneous shrinking of the basin of attraction, thereby making synchronization increasingly resistant to control. While a non-monotonic behavior could, in principle, arise if basin shrinkage was dominant, however this scenario has not been observed here.

{\rev 
Complementary to this analysis, we considered a second option to vary the intensity of the control: we varied the number of controlled nodes $M\le N$ and fixed $\mu$. This framework yields a more heterogeneous control and is closer to real-world applications where all nodes cannot always be controlled. Results show that the minimal number of nodes $M_c$ needed to desynchronized phases beyond $R_{\rm thr}$ increases with $K_2$ (\cref{fig:M_c}), similarly to $\mu_c$ (\cref{fig:mu_c}).
}

\subsection{Random Hypergraphs}\label{ssec:Exp1Random}

Let us now further validate these results with a more complex topology in system \eqref{eq:HOKM}. We here consider random hypergraphs obtained from the Erdős-Rényi algorithm~\cite{iacopini2019simplicial}. Given a set of $N$ oscillators, we build the hypergraph as follows: we add a 1-hyperedge (edge) for each pair of nodes with probability $p_1$ and a 2-hyperedge (triangle)  between each triplet with probability $p_2$. The mean degree and hyperdegree are then $\md1=Np_1$ and $\md2=N(N-1)p_2$. To compare the results with \cref{ssec:Exp1Ring}, we set $\md1=2r$ and $\md2=2r(r-1)$, with $r=2$ and 3.

These random hypergraphs do not have rotational invariance, and hence twisted states are no longer supported, except for the synchronized state. In the uncontrolled system, $\mu=0$, previous research showed that for low $K_2$, full synchronization attracts most initial conditions~\cite{zhang2023deeper}. As $K_2$ is increased, 2-cluster states, and then disordered states, attract most initial conditions instead~\cite{zhang2023deeper}; in parallel, the linear stability of full synchronization also increases even though its attraction basin shrinks. 

\begin{figure}[t]
	\centering
	\includegraphics[width=\linewidth]{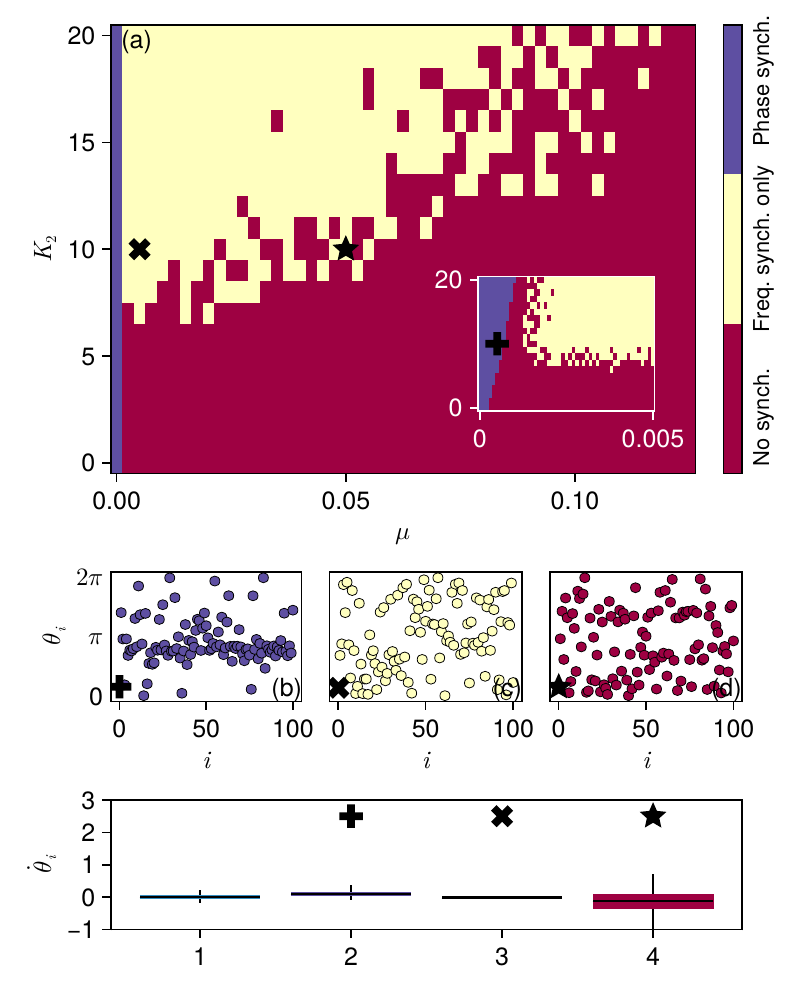}
	\caption{
		\textbf{Desynchronizing phase oscillators on random hypergraphs.}
		We show (a) the division in three zones: phases synchronization ($\hat{R}\geq 0.4$, blue), frequency synchronization only ($\hat{R}<0.4$ and $\hat{R}_{\dot\theta}\geq 0.5$, yellow), and no synchronization (otherwise, red). The inset displays details of the low $\mu$ region. Panels (b)-(d) show example snapshots of these regimes with {\rev $\mu = 0.0005$, $\mu=0.005$ and $\mu=0.05$}, respectively; (e) shows the boxplots of the associated effective frequency distributions, and that of the natural frequencies (left).
		Parameters are set to $N=100$, $K_1=1$, $\md1=2r$, $\md2=2r(r-1)$ and $r=2$.
	}
	\label{fig:K2_v_mu_r_2_random}
\end{figure}

Results for random hypergraphs are similar to those for hyperrings. \Cref{fig:K2_v_mu_r_2_random}a shows, for $r=2$, the phase diagram with the three regimes, obtained by combining the information from $\hat R$ and $R_{\dot\theta}$. It is clear that phase synchronization decreases as $\mu$ increases, while it increases as $K_2$ increases. Stated differently, the heatmap for $\hat R$ shows a gradient similar to hyperrings in the $(\mu, K_2)$ space (see also~\cref{fig:K2_v_mu_random_complement}a in Appendix~\ref{app:SuppFig}).
At lower $K_2<5$, the uncontrolled system remains fully synchronized (blue), but the control achieves incoherence (red) for $\mu>0$. 
As the triadic coupling strength $K_2$ is increased, the value of $\mu$ required to reach incoherence is larger, and intermediate values desynchronize phases but not the frequencies (yellow). 
Interestingly, zooming into the region associated to small values of $\mu$ (inset) reveals a band of no synchronization (red) at values of $\mu$ smaller than the region of frequency synchronization (yellow).  
The fully synchronized (\cref{fig:K2_v_mu_random_complement}b) and fully incoherent states (\cref{fig:K2_v_mu_random_complement}d) are similar to those observed for hyperrings. However, the phase desynchronized ones (\cref{fig:K2_v_mu_random_complement}c) are different because hyperrings are much more symmetric than random hypergraphs. They can be distinguished from full incoherence because of the spread of their effective frequency distributions (\cref{fig:K2_v_mu_random_complement}e).
Again, we observe that the case $r=3$ exhibits qualitatively equivalent results (\cref{fig:k2_v_mu_r=3_Random} Appendix~\ref{app:SuppFig}).

%\cref{fig:mu_c_random} 
%are analogous to \cref{fig:K2_v_mu,fig:mu_c} in the hyperring case, and concern $r=2$. 
%The phase diagram showing the three regimes in parameter space is shown in \cref{fig:K2_v_mu_r_2_random} for $r=2$

\begin{figure}[t]
    \centering
    \includegraphics[width=\linewidth]{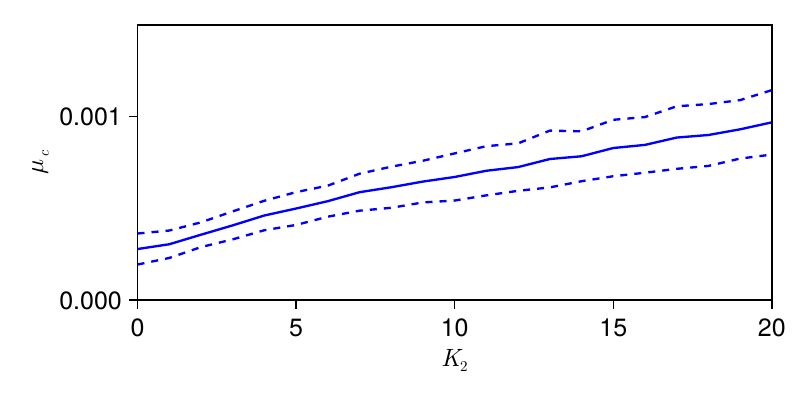}
    \caption{
    \textbf{Minimum control strength $\mu_c$ required to desynchronize $\hat R$ < 0.4 increases with triadic coupling strength $K_2$ on random hypergraphs.} 
    We show $\mu_c$ averaged (solid line) over 50 frequency distributions $\bm{\omega}$, and one standard deviation away (dashed lines).
    Parameters are, as in \cref{fig:K2_v_mu_r_2_random}, set to $N=100$, $K_1=1$, $\md1=2r$, $\md2=2r(r-1)$ and $r=2$.}
    \label{fig:mu_c_random}
\end{figure}

%Here, we observe again that increasing $K_2$ strengthens the local stability of phase synchronization and, thus, it is an obstacle to the desynchronization. The above-defined quantity $\mu_c$ increases and the phase-synchrony zone in \cref{fig:K2_v_mu_r_2_random}(a) enlarges with $K_2$. 

Similarly to the case of hyperrings, the minimal control strength $\mu_c$ required to reach a given level of phase desynchronization increases with triadic coupling strength $K_2$ ({\rev and \cref{fig:mu_c_random,fig:K2_v_mu_varying_R_thr_random} in \cref{app:SuppFig}}).

In summary, stronger control is required to desynchronize oscillators---in phase or in frequencies---as higher-order coupling strength is increased, once the system is initialized from fully synchronized initial conditions. Thus, we only observe the influence of the increasing linear stability.
These observations are consistent with those from Sec.~\ref{ssec:Exp1Ring}. To be able to observe the effect of the shrinking of the basin of attraction, we have to move the initial condition away from the fully synchronized state. 

{\rev
% \subsection{Analysis of the required number of controlled nodes}

Similarly to the case of hyperrings, we also varied the number of controlled nodes $M$ with fixed $\mu$. This slight variation results in a more heterogeneous control action.  
Contrary to hyperrings, here $M_c$---the minimal number  to desynchronize the phases of the system---changes  non-monotonically with $K_2$~(\cref{fig:M_c_random}). This is related to shrinking of the basins, and we investigate the effect of varying $M$ more systematically in the next section. %The difference in the results between the two topologies when we vary $M$ should be attributed to the breaking of symmetry in the structure.
}

\section{The case of perturbed synchronization as initial conditions}\label{sec:Exp2}

\begin{figure*}[htb]
    \centering
    \includegraphics[width=\linewidth]{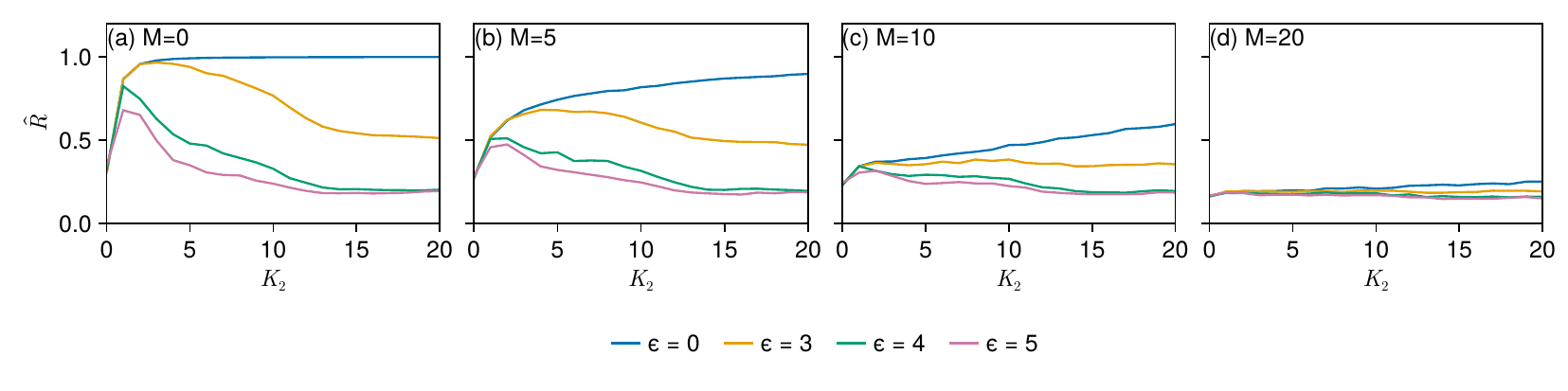}
    \caption{
    \textbf{Desynchronizing phase oscillators on a hyperring, with partial control.}
    We show the level of phase synchronization $\hat R$ against triadic coupling strength $K_2$, by controlling (a)-(d) $M=0$, $5$, $10$, and $20$ nodes, respectively.
    For each value of $M$, we show three values of initial distance to basin center $\epsilon=0$, 3, and 5.
    Other parameters are set to $N=100$, $r=2$ and $K_1=1$. 
    Order parameter $\hat R$ is averaged over 10 random realizations of frequencies $\bm{\omega}$ and 50 of initial conditions $\theta_0$.
    }
    \label{fig:HatR_r_2_K1_1}
\end{figure*}

In \cref{sec:Exp1}, we observed how the control affects
%the effect of the increasing of the linear stability. The reason is that we have set initial conditions to 
the synchronized state once the system is initialized close to the latter. However, initial conditions can, in general, be arbitrary and not directly related to the synchronous solution.
As the control can be seen as a force that tries to push the trajectories out of the basin of attraction of the synchronous state, the actual position of the initial condition plays a crucial role. 
It is easier for the system to avoid synchronization if the initial conditions are close to the border of the basin, whereas it should be more difficult if one starts close to its center. 
In this respect, it is important to examine how the system behaves by varying the initial conditions. 
Additionally, we want to compare control schemes where the number of pinned nodes, $M < N$, is varied, which may be easier and less costly to apply in practice.

%setting $M=N$ is very restrictive from a pinning control perspective and letting $M$ vary in smaller values is more realistic.

We consider system \eqref{eq:HOKM} and investigate how the level of phase synchronization $\hat R$ varies with $K_2$, $M$ and the distance between the initial conditions and the phase synchronized state of the system. 
%The larger is the latter, the less the linear stability of the full synchronization plays a role.
We modulate this distance with a parameter $\epsilon \ge 0$ as follows. First, we let the system converge to the phase-synchronized state $\theta_{\rm fs}$ as in \cref{sec:Exp1}. Second, we perturb it as
\begin{equation}
    \theta_0 = \theta_{\rm fs} + \epsilon u\, ,
\end{equation}
where $u\sim U[-\frac12,\frac12]^N$. 
Parameter $\epsilon$ controls the initial expected distance from full synchronization. If it is small, the initial conditions tend to lie close to the center of the basin. 
Note, however, that the basin does not necessarily need to be ball-shaped~\cite{Zhang2021BasinsWithTentacles}.
% maybe just say something about the potentially non ball shape
%It has to be noted that, even though we can estimate numerically the relative size of the basin of attraction as made in~\cite{zhang2023deeper}, its actual shape remains unknown. It is not guaranteed that it has a ball shape, neither that $\theta_{\rm fs}$ can be considered as its center. However, we argue that this reasoning provides a good proxy of the actual depth.
We fix {\rev $\mu = 0.5$}, and the intensity of the control is now solely controlled by the number of pinned nodes, $M$.

As in \cref{sec:Exp1}, numerical examples are shown for $N = 100$, {\rev $t_{tr}=250$, $t_i = t_{tr} + 190$, $t_f = t_{tr} + 200$,} $K_1=1$, {\rev $\sigma=0.1$,} and for both hyperring and Random Erd\H{o}s-Renyi structures parametrized by $r = 2$ or 3. Parameter $K_2$ is varied between $0$ to $20$ and $\epsilon$ between $0$ and $2\pi$. For each value of $(K_2,\epsilon)$, we performed $50$ random realizations of \eqref{eq:HOKM} where $u$ is sampled as shown above and the $M$ pinned nodes are distributed uniformly at random among the hypergraph nodes.

\subsection{Hyperrings}\label{ssec:Exp2Rings}

Let us first consider the same hyperring structure as in \eqref{eq:HOKMRing}. In \cref{fig:HatR_r_2_K1_1}, we report the results we obtained for $\hat R$, where we fixed $N = 100$, $r = 2$, and $K_1 = 1$ and made $\epsilon$ and $K_2$ vary. 
%We also sampled several $\omega$ and $u$ distributions. 
Further results for $r=3$ are shown in~\cref{fig:HatR_r_3_K1_1} in \cref{app:SuppFig}, which appear to be qualitatively equivalent.

First, for $M = 0$ (\cref{fig:HatR_r_2_K1_1}(a)), i.e., the uncontrolled system, and $\epsilon=0$, i.e., $\theta_0$ is the "center" of the basin, we observe that increasing $K_2$ leads to an increase in $\hat{R}$. This aligns with the results shown in Ref.~[\onlinecite{zhang2023deeper}], thus indicating deeper synchronization basins. 

By increasing $\epsilon$, a non-monotonic behavior emerges as $\hat{R}$ initially increases with $K_2$ and then decreases. This again aligns with previous work~\cite{muolo2025when}, showing that moderate higher-order interactions enhance synchronization, while strong ones hinder synchronization by reducing the basin of attraction. Indeed, as $\epsilon$ grows, $\theta_0$ is more likely to lie out of it and then the system goes disordered. 

When control is applied with $M$ taking intermediate values, e.g., $5$ or $10$, (\cref{fig:HatR_r_2_K1_1}(b)-(c)) and $\epsilon=0$, then $\hat R$ monotonically increases with $K_2$. In this regime, the control effectively reduces synchronization, although the presence of higher-order interactions acts as an obstacle. This is the analogous result of \cref{ssec:Exp1Ring}, that is higher-order interactions hinder the control when the system starts near the fully synchronized state. 

For the choices $\epsilon=3$, $4$, or $5$, we recover the same type of non-monotonic behavior as above, although the overall $\hat{R}$ values are lower due to the action of the control. Beyond a certain threshold, $\hat{R}$ begins to decrease with increasing $K_2$, and the curves tend to coincide with the uncontrolled case.

For $M = 20$ and whatever $\epsilon$ value, the $\hat{R}$ curves become nearly constant. Hence, the number of controllers is sufficient to achieve full {\rev phase} desynchronization regardless of the value of $K_2$. In other words, even when only a fraction of the nodes are pinned, the minimally-invasive Hamiltonian control can desynchronize a higher-order Kuramoto system.

It is also worth commenting on the apparent magnitude of $\epsilon$ which might seem large at first glance as the role of $\epsilon$ is to implement a relatively small deviation from the initial condition. However, the deviation defined as $\epsilon u$ means that the initial condition is drawn from a hypercube centered on the fully synchronized state $\theta_{\rm fs}$. The relative volume of this cube is actually $(\frac{\epsilon}{2\pi})^N$, which is of the order of $10^{-10}$ when $\epsilon = 5$. Thus, the perturbation remains small compared with the size of the full state space. 
%{\color{green}[Ric: is "phase space" correct here?]}

From these results, we can draw the following conclusions. First, when the system is initialized deep within the basin of attraction and $K_2$ is not too large, by increasing $K_2$ increases the strength of the control required for {\rev phase} desynchronization. This observation is consistent with the findings reported in~\cite{zhang2023deeper,von2024higher,Moriame2025Hamiltonian}. Second, when the system is initialized far from full synchrony and $K_2$ is large, the uncontrolled system does not synchronize, and control primarily serves to accelerate desynchronization and achieve lower $\hat{R}$ values. 
%Finally, we observe no qualitative difference between the controlled and uncontrolled dynamics. The effect of control appears to be a mere downward translation of the $\hat{R}$ curves.

\subsection{Random hypergraphs} \label{ssec:Exp2Random}

\begin{figure*}
    \centering
    \includegraphics[width=0.99\linewidth]{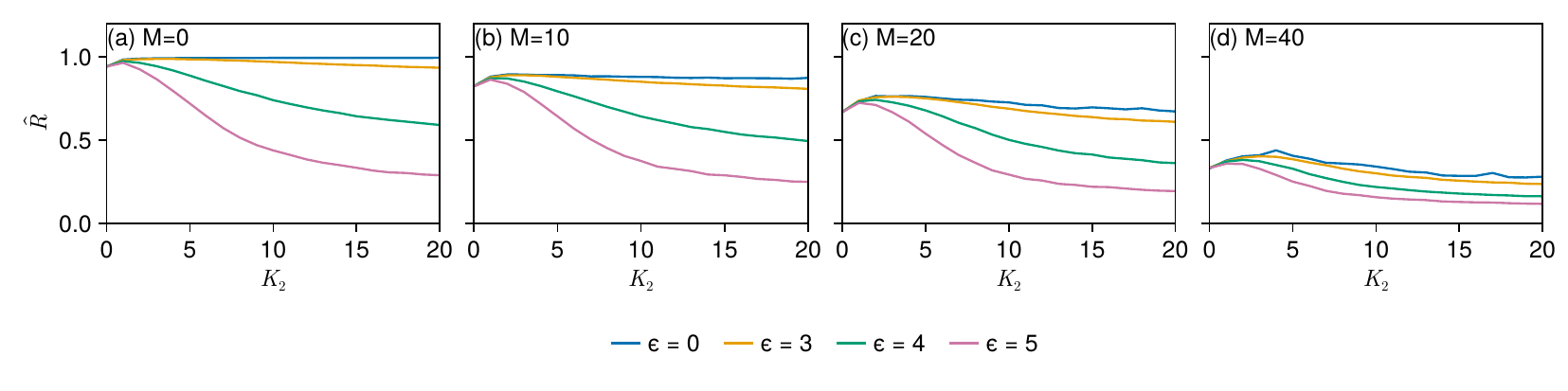}
    \caption{
    %Averaged $\hat R$ values, over 50 $\bm{\omega}$ distributions, each associated with a realization of the Erd\H{o}-Rényi hypergraph generation, in function of $\epsilon$, $M$ and $K_2$. Here we set $N=100$, and $K_1=1$, and $r=2$ so that $p_1=4/N$ and $p_2=4/N^2$. The panels (a) to (c) display respectively $M=0$, $M=20$ and $M=40$.
    \textbf{Desynchronizing phase oscillators on a random hypergraph, with partial control.}
    We show the level of phase synchronization $\hat R$ against triadic coupling strength $K_2$, by controlling (a)-(c) $M=0$, $20$, and $40$ nodes, respectively.
    For each value of $M$, we show values of initial distance to basin center $\epsilon$ between 0 and $2\pi$.
    Other parameters are set to $N=100$, $K_1=1$, $p_1=4/N$ and $p_2=4/N^2$. 
    Order parameter $\hat R$ is averaged over 10 random realizations of frequencies $\bm{\omega}$ and 50 of initial conditions $\theta_0$.
    }
    \label{fig:HatR_Random_r_2_K1_1}
\end{figure*}

Let us return to consider Random Erd\H{o}s-Rényi hypergraphs, as in \cref{ssec:Exp1Random}. The results we obtain, shown in \cref{fig:HatR_Random_r_2_K1_1}, are qualitatively similar to the case of hyperrings in \cref{ssec:Exp2Rings}. 

Without control, $M=0$, and with small values of $\epsilon$ ($0$ and $3$), the system remains synchronized regardless of the values of $K_2$ (\cref{fig:HatR_Random_r_2_K1_1}(a)). With larger $\epsilon$, the non-monotonic effect appears as $\hat R$ slightly increases and then decreases. The non-monotonic nature of the curves is less obvious than in \cref{ssec:Exp2Rings} because $\hat R$ is already close to $1$ for $K_2=0$.

With control, $M=10$, $20$, and $40$ nodes, the overall value of $\hat R$ decreases and all curves are pulled downward (\cref{fig:HatR_Random_r_2_K1_1}(b)-(c)). The non-monotonic behavior of the curve is also preserved, and is even more pronounced, as it is already present at the lowest $\epsilon$ values for $M=20$ and 40. 

These results are the direct consequences of the shrinking of the basin~\cite{zhang2023deeper,muolo2025when}. With moderate $K_2$ values, the synchronization is strengthen, but with {\rev large} $K_2$ the control achieves the {\rev phase} desynchronization more easily. This occurs because the control can push more easily the trajectories out of the basin, which has shrunk due to the strong higher-order coupling.

\section{Conclusion}
\label{sec:conclusion}

In this work, we implemented the desynchronizing pairwise minimally invasive Hamiltonian control~\cite{asllani2018minimally} on the higher-order Kuramoto model with pairwise and three-bodies interactions. We modified the control method so that it only requires knowledge of the parameters and states of the controlled nodes without having to know their connectivity. We used hypergraphs with hyperrings structures and then further validated our results on random Erd\H{o}s-Rényi hypergraphs. We measured the synchronization of the controlled system under several parameterizations to determine whether the higher-order interactions, quantified by the higher-order coupling strength, make the control more effective or not. In other words, we examined where and how the two antagonistic effects of higher-order interactions, i.e., the increasing of linear stability and narrowing of the basin of attraction of synchronization, influence the control efficiency.

We showed that the control was able to desynchronize the system, {\rev both in phases and frequencies,} provided that either a sufficiently large number of nodes is controlled or the control strength is sufficiently large. This is consistent with Ref.~[\onlinecite{Moriame2025Hamiltonian}], where the pairwise version of the control considered in that study was sufficient, in some cases, to desynchronize the higher-order network of oscillators. 
At the same time, our results go beyond those of Ref.~[\onlinecite{Moriame2025Hamiltonian}] in two respects. First, we consider a minimally invasive version of the control, which improves the previous computationally heavy control term \cite{Moriame2025Hamiltonian}. Second, in our case, the pairwise control achieves desynchronization even when all nodes are controlled, whereas this was only observed when pairwise interactions were dominant \cite{Moriame2025Hamiltonian}. This difference can be attributed to the topologies considered: all-to-all networks~\cite{Moriame2025Hamiltonian} make synchronization much more robust and, therefore, require a more general control strategy to guarantee desynchronization.

When the system starts with synchronized initial conditions, we observed that, by increasing the higher-order coupling strength, the control strength required to desynchronize the system increases. 
We observed that the control made the system to move from full synchronization to either frequency synchrony without phase synchrony, e.g., cluster or twisted states, or complete incoherence, with no phase or frequency synchrony.
The increase in $\mu_c$ with $K_2$ suggests that the effect of increased linear stability of full synchronization is dominant over the shrinking of the attraction basin when it comes to control desynchronization. 

%This occurs because of the induced effect of increasing linear stability. Even though this outcome is expected, it was not trivial that the shrinking of the basin would not play any role in this case. Its influence is, however, very limited, as it can only be observed with random hypergraphs, with large control strength and number of controlled nodes. Therefore, we can conclude that the increasing of the linear stability is the most influent aspect in this case.

In contrast, for other initial conditions, a non-monotonic behavior is observed. Intermediate values of the higher-order coupling strength make the control less effective, whereas larger coupling strengths enhance its efficiency. This behavior appears to reflect the effect of the progressive shrinking of the basin of attraction of the synchronized state. This non-monotonic behavior is in line with other recent studies~\cite{muolo2025when,wang2025moderate,skardal2025mixed}, where an analogous phenomenology was observed in different contexts. %In this work, we showed that this non-monotonic behavior also appears in a desynchronizing control context. 

%Moreover, the controlled system can exhibit this even though the uncontrolled system doesn't . {\color{green}[Ric: I'm not sure what this last sentence means.]} {\color{red} Martin: I reformulated, is it better?}

These results shed light on the importance of the dual effect of higher-order interactions, both on linear and basin stability, when control methods are involved.  In particular, the non-monotonic nature of this effect could be used for designing efficient control methods and appropriate interaction topologies. 

Future work could investigate, for example, the interplay between this non-trivial effect and other complex and more varied topologies, as well as the influence of the choice of controlled nodes, which are selected uniformly at random in this work. Building on recent results on the controllability of higher-order structures~\cite{chen2021controllability}, an interesting direction would be to increase the efficiency of the control by targeting specific nodes. Lastly, more application-oriented framework could directly consider non-reduced oscillatory models, as done, e.g., in Ref.~[\onlinecite{asllani2018minimally}] for the Stuart-Landau oscillator~\cite{nakao2014complex}, for which this study on the Kuramoto model paves the way.

%\MartinAdd{Other efforts should focus on the pinning control design to further improve it, in particular by investigating how to find the best pinned nodes subset given a fixed number $M$ so that the control reaches its maximal possible efficiency. Further more, although the Higher-order Kuramoto model is an abstract paradigmatic model of phase oscillators, we mentioned above can be obtained from a vaster diversity of limit cycles oscillators systems thanks to phase reduction reduction techniques. One example is the Stuart-Landau model with higher-order interactions~\cite{leon2024higher}. As initiated in Ref.~[\onlinecite{asllani2018minimally}], future efforts should be done to bridge between the controled Kuramoto system \eqref{eq:HOKM} and control of more complex systems like Stuart-Landau.}

\medskip

\noindent \textbf{Acknowledgments:} \\
\noindent R.M. acknowledges JSPS KAKENHI 24KF0211 for financial support. M.L. is a Postdoctoral Researcher of the Fonds de la Recherche Scientifique–FNRS. Part of the results were obtained using the computational resources provided by the ``Consortium des Equipements de Calcul Intensif" (CECI), funded by the Fonds de la Recherche Scientifique de Belgique (FRS-FNRS) under Grant No. 2.5020.11 and by the Walloon Region.

\noindent \textbf{Author contributions:} \\
M.M. and R.M. conceptualized the study. M.M. and M.L. developed the methodology. M.M. carried out the theoretical analysis and the simulations, validated the results, curated the visualization, and wrote the manuscript. T.C. and M.L. supervised the project. All authors discussed the results, reviewed, and edited the manuscript.

\noindent \textbf{Conflict of interest:} \\
The authors have no conflicts to disclose.

\noindent \textbf{Data availability:} \\
The data that support the findings of this study are available within the article.

\bibliography{bib}

\clearpage

\appendix
\onecolumngrid 

%\the\textwidth

\section{Computation of the control term}
\label{app:control}

The original pairwise Hamiltonian pinning control was developed in Ref.~[\onlinecite{gjata2017using}], based on Hamiltonian control theory~\cite{vittot2004perturbation,ciraolo2004control} and the Hamiltonian embedding of the Kuramoto model~\cite{witthaut2014kuramoto}. This method has then been perfected in two ways. 

In Ref.~[\onlinecite{asllani2018minimally}], the authors went a step further by reducing the control term's magnitude in order to make the method minimally invasive. In short, they isolated the dominant terms so that the control adapts its magnitude to grow with $K_1^2$ and $R(t)$. It is thus significant only when the system is (likely to be) synchronized, i.e., when $K_1$ and/or $R$ are large, but negligible otherwise. By doing so, the authors made the control method more likely to be implemented in a practical case, e.g, to prevent the emergence of seizures.

On the other hand, it was shown~\cite{Moriame2025Hamiltonian} that the (not minimally invasive) pairwise control can (partially) desynchronize the higher order system in several cases, namely if the pairwise interactions are strong and the higher order ones weak. In the opposite case, Ref.~[\onlinecite{Moriame2025Hamiltonian}] proposes to use the higher-order generalization of the Hamitonian control which can desynchronize the higher-order system in any case. However, it comes with the cost of an increasing number of high magnitude terms, which would make it even harder to implement in a practical context.

In this work, we propose to use the minimally invasive version of the pairwise control in order to desynchronize higher-order systems of Kuramoto oscillators. By doing so, we accept that in some cases the control will not achieve its goal, but, on the other hand, we can understand more precisely the cases where this minimal control action is sufficient and when it should be reinforced with higher-order terms.

We nevertheless introduce some slight modifications from the method used in Ref.~[\onlinecite{asllani2018minimally}]. (i) Whereas the latter requires the knowledge of the pairwise topology of the system, which is a real constraint in a practical context, we relax this by computing the control term as if the pinned nodes were connected by a clique. It makes the control both easier to compute and potentially more powerful (see bellow). (ii) We restrict the computation of the control term to the very $M$ pinned nodes. The control acts thus as a proper feedback loop: from the measure of the phases and natural frequencies of the pinned nodes only, the control term is computed and then injected into those nodes' dynamics without requiring the knowledge of the global order parameter like in Ref.~[\onlinecite{asllani2018minimally}].

Let us now develop how the control term $p_i$ of \cref{eq:HOKM} is defined, by following the same methods of Refs~[\onlinecite{gjata2017using,asllani2018minimally}] but taking into account the little modifications (i) and (ii) we have just mentioned. The idea of Ref.~[\onlinecite{gjata2017using}] consists in adding, to a subset of $M$ nodes, an additive term
\begin{equation}
    h_i=\left\{
    \begin{matrix}
        -\frac12\frac{\partial \{\Gamma V,V\}}{\partial I_i}_{\text{\Large|} T_{1/2}}&\text{ if } i=1,\dots,M, \\
        0&\text{ otherwise,}
    \end{matrix}
    \right.
\end{equation}
where 
\begin{equation}\displaystyle
    H({\bm{I}}, {\bm{\theta}})= H_0({\bm{I}}) + V({\bm{I}}, {\bm{\theta}}) =\sum_iI_i\omega_i - \frac{K}{ \langle k^{(1)}\rangle}\sum_{ij}A_{ij}\sqrt{I_iI_j}(I_j-I_i)\sin(\theta_j-\theta_i)
\end{equation}
is a Hamiltonian function of action and angles variables $I_i$ and $\theta_i$. It is such that the resulting Hamiltonian system embeds the classical Kuramoto Model on the invariant torus $T_{1/2}=\{({I},{\theta})|\forall i : I_i=1/2\}$. This function exists in higher-order versions but we here restrict ourselves to the pairwise case. Finally, $\Gamma$ is the pseudoinverse operator of $H_0$.

The term $h_i$ actually uses all the network parameters, i.e., $K_1$, $N$, the entries $A_{ij}$, the frequencies $\omega_i$ and the phases $\theta_i$ (such that the mentioned indexes $i,j\in 1,\dots,M$) and is computed (for $i=1,\dots,M$) as
\begin{eqnarray}\displaystyle
    \label{eq:FullControlTerm2D}
    h_i&=&\frac{1}{2}\left(\frac{K_1}{ \langle k^{(1)}\rangle}\sum_{k=1}^MA_{ki}\cos(\theta_k-\theta_i)  \right) \times \left(-\frac{K_1}{ \langle k^{(1)}\rangle}\sum_{k=1}^MA_{ki}\frac{\cos(\theta_k-\theta_i)}{\omega_k-\omega_i} \right)\\
    &-&\frac{1}{2}
    \left(-\frac{K_1}{ \langle k^{(1)}\rangle}\sum_{k=1}^MA_{ki}\sin(\theta_k-\theta_i) \right) \times \left(-\frac{K_1}{ \langle k^{(1)}\rangle}\sum_{k=1}^MA_{ki}\frac{\sin(\theta_k-\theta_i)}{\omega_k-\omega_i} \right) \nonumber \\
    &+&\frac{1}{2}\sum_{j=1}^M\left\{\left(-
    \frac{K_1}{ \langle k^{(1)}\rangle}A_{ij}\cos(\theta_j-\theta_i) \right)\times\left(-\frac{K_1}{ \langle k^{(1)}\rangle}\sum_{k=1}^M A_{kj}\frac{\cos(\theta_k-\theta_j)}{\omega_k-\omega_j}  \right) \right. \nonumber \\
    &&\left.-\left(-\frac{K_1}{ \langle k^{(1)}\rangle}\sum_{k=1}^MA_{jk}\sin(\theta_k-\theta_j)\right)\times\left(\frac{K_1}{ \langle k^{(1)}\rangle}A_{ij}\frac{\sin(\theta_j-\theta_i)}{\omega_j-\omega_i} \right) \right\} \nonumber .
\end{eqnarray}

As said above (i), we here propose compute the pinning control term in the same way that if $\bm{ A}$ actually defined a $M-$clique connecting the $M$ pinned nodes all together, in other words
\begin{eqnarray}\displaystyle
    \hat{h}_i&:=&\frac{1}{2}\left(\frac{K_1}{ \langle k^{(1)}\rangle}\sum_{k=1}^M\cos(\theta_k-\theta_i)  \right) \times \left(-\frac{K_1}{ \langle k^{(1)}\rangle}\sum_{k=1}^M\frac{\cos(\theta_k-\theta_i)}{\omega_k-\omega_i} \right)\\
    &-&\frac{1}{2}
    \left(-\frac{K_1}{ \langle k^{(1)}\rangle}\sum_{k=1}^M\sin(\theta_k-\theta_i) \right) \times \left(-\frac{K_1}{ \langle k^{(1)}\rangle}\sum_{k=1}^M\frac{\sin(\theta_k-\theta_i)}{\omega_k-\omega_i} \right) \nonumber \\
    &+&\frac{1}{2}\sum_{j=1}^M\left\{\left(-
    \frac{K_1}{ \langle k^{(1)}\rangle}\cos(\theta_j-\theta_i) \right)\times\left(-\frac{K_1}{ \langle k^{(1)}\rangle}\sum_{k=1}^M \frac{\cos(\theta_k-\theta_j)}{\omega_k-\omega_j}  \right) \right. \nonumber \\
    &&\left.-\left(-\frac{K_1}{ \langle k^{(1)}\rangle}\sum_{k=1}^M\sin(\theta_k-\theta_j)\right)\times\left(\frac{K_1}{ \langle k^{(1)}\rangle}\frac{\sin(\theta_j-\theta_i)}{\omega_j-\omega_i} \right) \right\} \nonumber .
\end{eqnarray}
This approach has the double advantage of freeing us from the need to know the actual connections $A_{ij}$ between the pinned nodes and of using the control at the maximum of its potential power, as all the possible terms are considered in the sums (no $A_{ij}=0$ as is \eqref{eq:FullControlTerm2D}).

Then, we can write~\cite{asllani2018minimally}
\begin{equation}\label{eq:hath}
    \hat{h}_i = -\frac12 K_1^2\frac{M^2}{\langle k^{(1)}\rangle^2}R_M\hat{R}_{M,i}\cos(\Psi_M-\hat{\Psi}_{M,i}) + \mathcal{B}_i\, ,
\end{equation}
where $\mathcal{B}_i$ is a term that, in general, can be neglected~\cite{asllani2018minimally}. The latter can thus be removed without losing the control efficiency. The final resulting pinning control term, that is minimally invasive~\cite{asllani2018minimally}, is the term \eqref{eq:control}. Note that here, $R_M$ is used in \eqref{eq:hath} instead of $R$ like in  Ref.~[\onlinecite{asllani2018minimally}], so that the control term computation do not require the knowledge of any feature of the uncontrolled nodes (ii).
%\clearpage

\section{Supporting numerical results}\label{app:SuppFig}

This Section displays the Figures which complement the numerical results of \cref{sec:Exp1,sec:Exp2}. Its structure is set in parallel with those Sections.

\subsection{The case of synchronized initial conditions}

Let us first introduce the notion of $n^{th}$ order parameter\cite{gong2019lowdimensional}
\begin{equation}\label{eq:nthOrderParam}
    R_{(n)}e^{\imath \Psi_{(n)}} = \frac{1}{N} \sum_{j=1}^N e^{n\imath\theta_j}\, ,
\end{equation}
which are used to detect cases of clustered synchronization. Namely, $R_{(n)}\in[0,1]$ and 
\begin{equation*}
    R_{(n)}(t)=1 \, \Longleftrightarrow \, \forall\, i=1,\dots,N: \theta_i\in \left\{\theta_1(t) + \frac{2\pi j}{n}\, |\, j=1,\dots,n \right\}\,.
\end{equation*}
In other words, $R_{(n)}=1$ when the phases are distributed into $n$ groups of equally-valued members that are separated by an angle of $\frac{2\pi}{n}$. If the phases are not perfectly $n-$clustered, i.e., the members of a cluster have close phases but not exactly equal and/or the distance between the clusters are not exactly $\frac{2\pi}{n}$, then $0\ll R_{(n)}<1\,$. Note however that the groups do not necessarily have the same size. In the same way as in \eqref{eq:AsymOrederPamram}, we define
\begin{equation}
    \hat R_{(n)} := \langle R_{(n)}(t) \rangle_{t \in [t_i,t_f]}\, 
\end{equation}
to measure the presence of cluster synchronization.

\begin{figure}[t]
    \centering
    \includegraphics[width=0.66\linewidth]{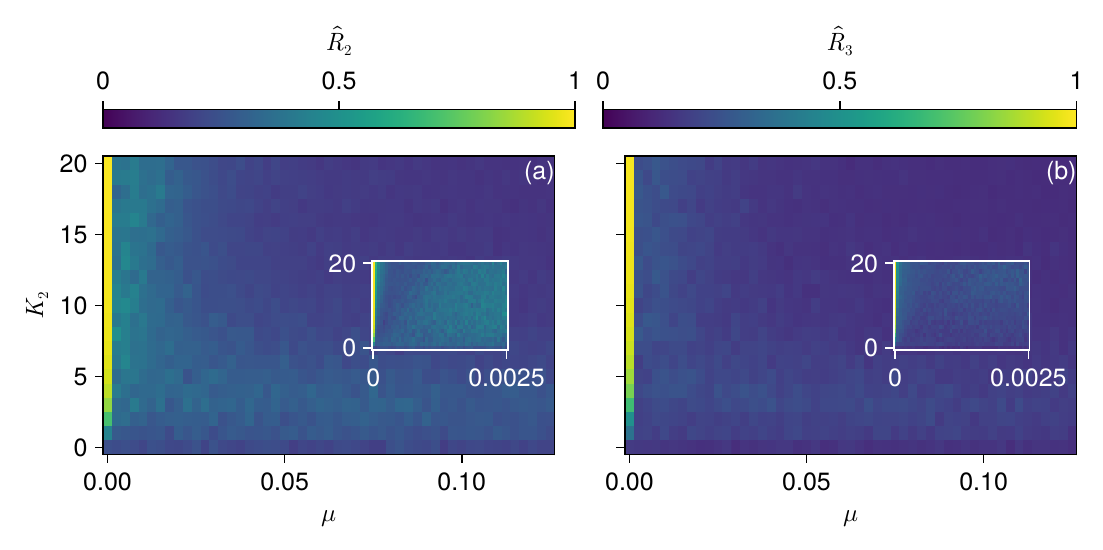}
    \caption{\textbf{Extension of \cref{fig:K2_v_mu}} encoding $\hat R_2$ (panel (a)) and $\hat R_3$ (panel (b)), the order parameters detecting resp. 2- and 3-cluster synchronization, as function of $K_2$ and $\mu$. We use a hyperring structure with $r=2$. As in \cref{fig:K2_v_mu}, each point ($K_2$, $\mu$) is the value averaged over 50 frequency distributions $\bm{\omega} \sim \mathcal{N}(0,\sigma^2)$, $N=100$, $K_1=1$, $\bm{\omega}\sim\mathcal{N}(0,\sigma^2)$ and $\sigma=0.1$.}
    \label{fig:K2_v_mu_R2_R3}
\end{figure}

Respectively in \cref{fig:K2_v_mu_R2_R3}(a) and \cref{fig:K2_v_mu_R2_R3}(b) we report the values of $\hat{R}_{(2)}$ and $\hat{R}_{(3)}$ obtained from the same trajectories that in \cref{fig:K2_v_mu}. We can first remark that for $\mu=0$ the values of $\hat{R}_{(2)}$ and $\hat{R}_{(3)}$ are very close to $\hat R$ observed in \cref{fig:K2_v_mu}(a), i.e., they rapidly grow with $K_2$ and are very close to 1 if $K_2\geq3$. This is expected, as a straightforward computation can show that $R_{(1)}=1 \Rightarrow R_{(2)}=1$ and $R_{(1)}=1 \Rightarrow R_{(2)}=1$, although the contrapositive is not true. With $\mu>0$, i.e., control activated, $\hat{R}_{(2)}$ and $\hat{R}_{(3)}$ are in general quite low, meaning that there are no notable presence of clustered synchronization. However, one can notice that $\hat{R}_{(2)}$ reaches intermediate values around 0.5 in a region that corresponds to the "frequency synchronization only" zone in \cref{fig:K2_v_mu}(b-c). This is in line with the partially-clustered trajectory that we observe in \cref{fig:K2_v_mu}(e), that looks like a $3-$cluster at first glance but is actually closer to a $2-$cluster because of the angular nature of the data (the two first groups we see on the figure actually form one single group, as their members are located close to 0 and $2\pi$, respectively).

\begin{figure}[ht]
    \centering
    \includegraphics[width=\linewidth]{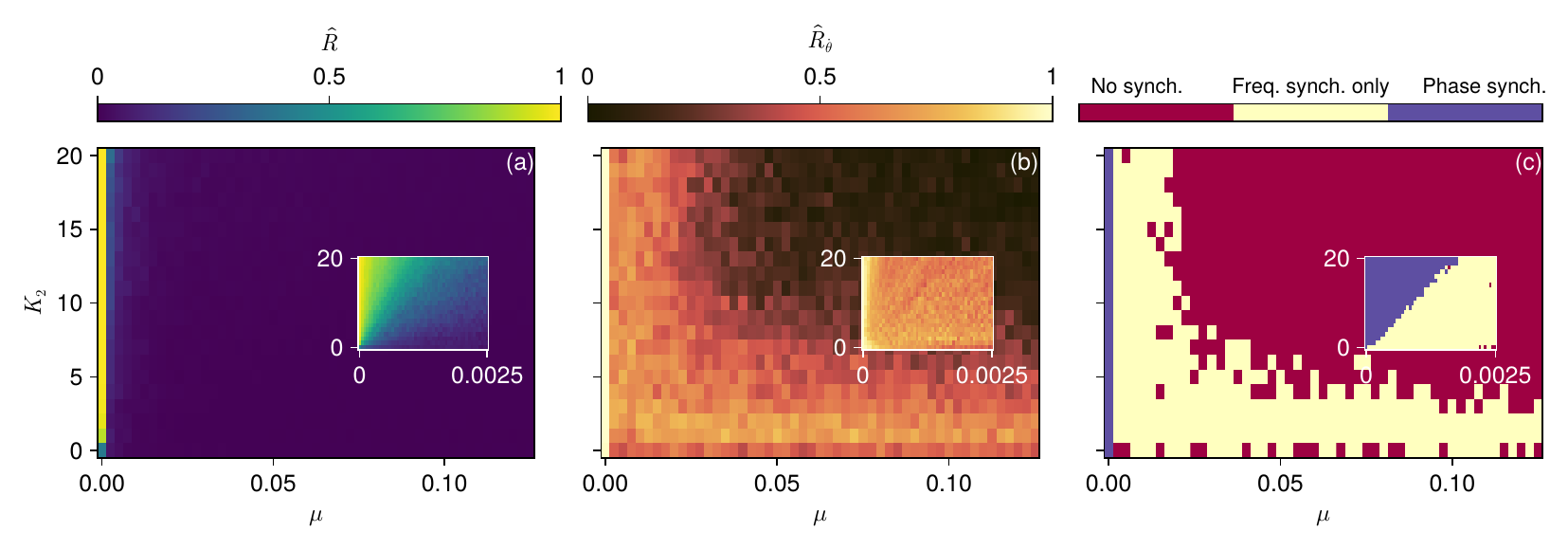}
    \caption{{\bf Desynchronizing phase oscillators on a hyperring with $r=3$.}
    Similarly to  \cref{fig:K2_v_mu}, we show (a) the level of phase synchronization $\hat R$, (b) of frequency synchronization $\hat{R}_{\dot\theta}$ and (c) the three regimes of the controlled system---phase synchronization (blue, $\hat{R}\geq 0.4$), only frequency synchronization (yellow, $\hat{R}<0.4$ and $\hat{R}_{\dot\theta}\geq 0.5$), and no synchronization (red, otherwise)---, all as function of $K_2$ and $\mu$. 
    For each value of ($K_2$, $\mu$), we show the value averaged over 50 frequency distributions $\bm{\omega} \sim \mathcal{N}(0,\sigma^2)$.
    Parameters are set to $N=100$, $K_1=1$, $\sigma=0.1$ and $r=3$.}
    \label{fig:k2_v_mu_r=3}
\end{figure}

\cref{fig:k2_v_mu_r=3} shows the results of the same experiment on hyperrings with $r=3$ and is analogous to \cref{fig:K2_v_mu} (case $r=2$). We can observe qualitatively equivalent results as in the case $r=2$, as the $\hat R\geq R_{\rm thr.}$ and $\hat R_{\dot\theta}\geq 0.5$ have the same shape. The main difference is that those zones, i.e., the phase synchrony and frequency synchrony zones, are wider than in the $r=2$ case. Indeed, with $r=3$ the links and triangles are denser in the hypergraph, i.e. there are more interactions between the nodes, which causes an increased stability of the synchronized state. The control goal is therefore harder to achieve.

\begin{figure}[ht]
    \centering
    \includegraphics[width=0.5\linewidth]{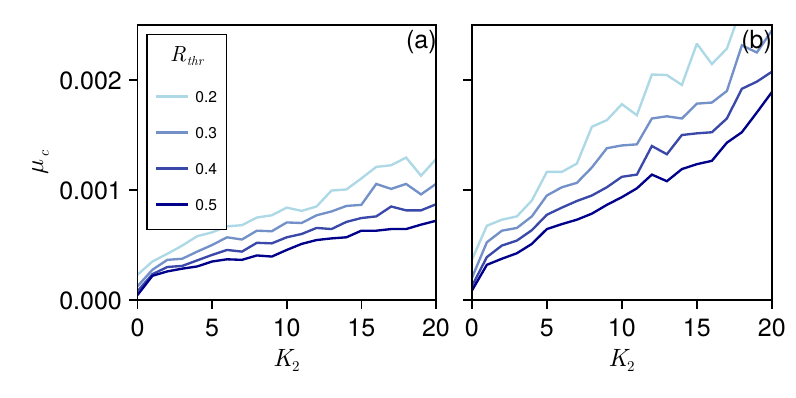}
    \caption{\rev \textbf{Extension of \cref{fig:mu_c}} encoding the minimal control strength $\mu_c$ required to desynchronize $\hat R < R_{\rm thr}$ on a hyperring, where $R_{\rm thr}$ varies in $[0.2,0.5]$.
    We show $\mu_c$ averaged over 50 frequency distributions $\bm{\omega}$. Panel (a): $r=2$; panel (b): $r=3$.
    The other parameters are set to $N=100$ and $K_1=1$.}
    \label{fig:K2_v_mu_varying_R_thr}
\end{figure}

{\rev In \cref{fig:K2_v_mu_varying_R_thr}, we show $\mu_c$ as a function of $K_2$ by varying the desynchronization threshold value $R_{\rm thr}$ on the hyperring topology. Independently of the value of $R_{\rm thr}$, the relation between $\mu_c$ and $K_2$ remains monotonous. This strengthens the observations made in \cref{fig:K2_v_mu} of \cref{ssec:Exp1Ring} and further confirms that they are independent of the threshold value, the latter being arbitrarily fixed. The results are equivalent for $r=2$ and $r=3$.}

\begin{figure}[ht]
    \centering
    \includegraphics[width=\linewidth]{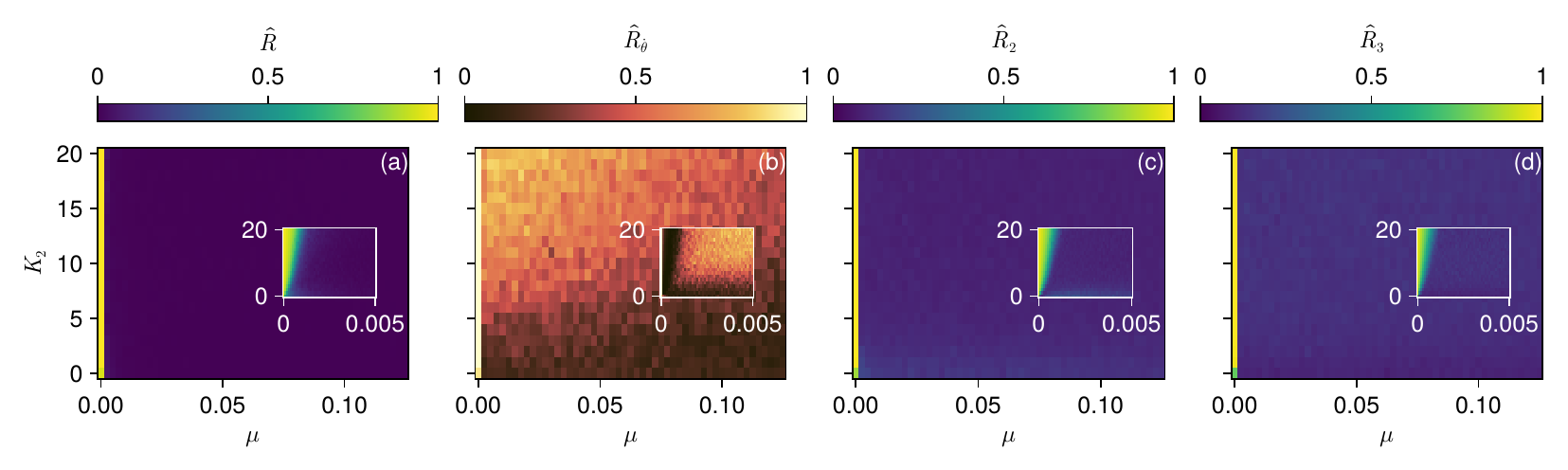}
    \caption{\textbf{Extension of \cref{fig:K2_v_mu_r_2_random},} displaying (a) $\hat R$, (b) $\hat R_{\dot\theta}$, (c) $\hat R_{(2)}$ and (d) $\hat R_{(3)}$, all as a function of $K_2$ and $\mu$. As in \cref{fig:K2_v_mu_r_2_random}, the trajectories are drawn from random hypergraphs with parameters set as $N=100$, $K_1=1$, $\hat R_{(2)}$, $\sigma=0.1$, $\md1=2r$, $\md2=2r(r-1)$ and for each value of ($K_2$, $\mu$), we show the value averaged over 50 frequency distributions $\bm{\omega} \sim \mathcal{N}(0,\sigma^2)$.}
    \label{fig:K2_v_mu_random_complement}
\end{figure}

 In \cref{fig:K2_v_mu_random_complement} we show additional panels related to \cref{fig:K2_v_mu_r_2_random}. Panels (a-b) exhibit resp. the $\hat R$ and $\hat R_{\dot\theta}$ data that correspond to the aggregated panel \cref{fig:K2_v_mu_r_2_random}(a). We observe no presence of cluster synchronization in this case.

\begin{figure}[ht]
    \centering
    \includegraphics[width=\linewidth]{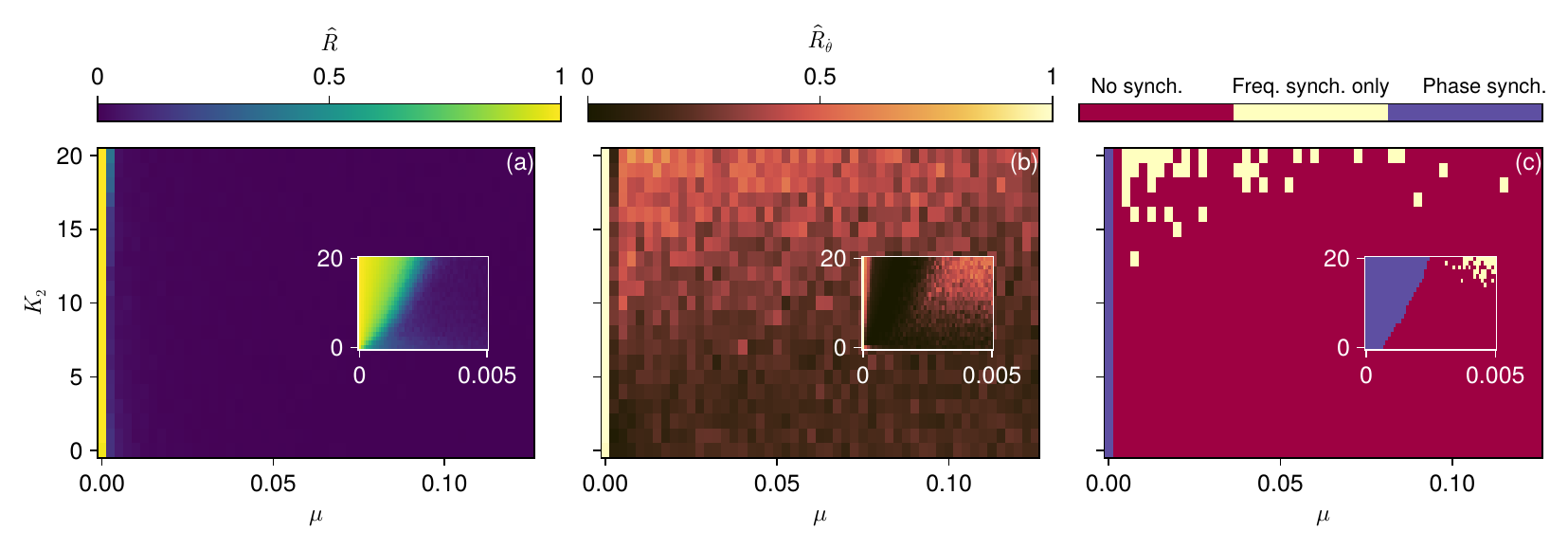}
    \caption{{\bf Desynchronizing phase oscillators on a random hypergraphs with $r=3$.}
    Similarly to  \cref{fig:K2_v_mu_r_2_random}, we show (a) the level of phase synchronization $\hat R$, (b) of frequency synchronization $\hat{R}_{\dot\theta}$ and (c) the three regimes of the controlled system---phase synchronization (blue, $\hat{R}\geq 0.4$), only frequency synchronization (yellow, $\hat{R}<0.4$ and $\hat{R}_{\dot\theta}\geq 0.5$), and no synchronization (red, otherwise)---, all as function of $K_2$ and $\mu$. 
    For each value of ($K_2$, $\mu$), we show the value averaged over 50 frequency distributions $\bm{\omega} \sim \mathcal{N}(0,\sigma^2)$ together with 50 realizations of random hypergraphs.
    Parameters are set to $N=100$, $K_1=1$, $\sigma=0.1$ and $r=3$.}
    \label{fig:k2_v_mu_r=3_Random}
\end{figure}

\cref{fig:k2_v_mu_r=3_Random} displays the analogous results for random hypergraph structures with $r=3$. We can observe that the results are in general qualitatively equivalent to the case $r=2$ in~\cref{fig:K2_v_mu_r_2_random}. The two main differences are that the phase synchrony region is larger while the region with only frequency synchronization is narrower. This may be explained by the increasing of both pairwise and higher-order interactions, which increase the local stability of phase synchronization and make the control goal harder to achieve. Moreover, other phase locking states are also harder to attain because of the absence of symmetry in the topology.% \MM{ceci est une ébauche d'expliquation et pourrait sûrement être amélioré, voire enlevé si on n'a rien de satisfaisant. [Ric: I modified it a bit. It's fine for me]}

\begin{figure}[ht]
    \centering
    \includegraphics[width=0.5\linewidth]{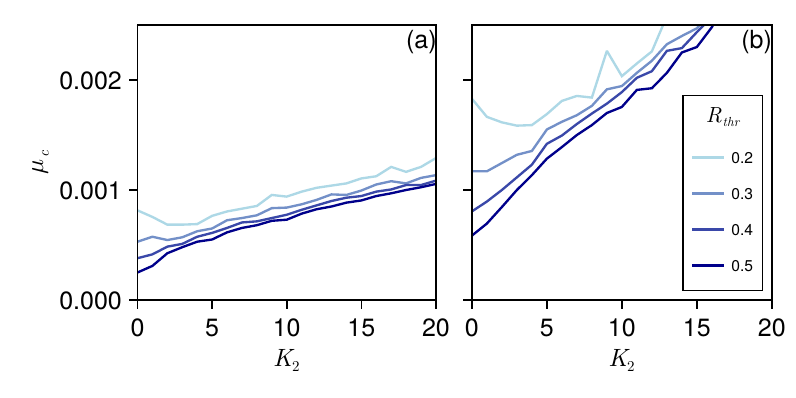}
    \caption{\rev \textbf{Extension of \cref{fig:mu_c_random}} encoding the minimal control strength $\mu_c$ required to desynchronize phases below $\hat R < R_{\rm thr}$ on random hypergraphs topologies, where $R_{\rm thr}$ varies in $[0.2,0.5]$.
    We show $\mu_c$ averaged over 50 random hypergraph topologies with $\md1 = 2r$ and $\md2 = 2r(r-1)$ and 50 frequency distributions $\bm{\omega}$, with (a) $r=2$ and (b) $r=3$.
    The remaining parameters are $N=100$ and $K_1=1$.}
    \label{fig:K2_v_mu_varying_R_thr_random}
\end{figure}

{\rev In \cref{fig:K2_v_mu_varying_R_thr_random}, we depict $\mu_c$ as a function of $K_2$ for a range of desynchronization threshold values $R_{\rm thr}$ on random hypergraph topologies. Analogously to the hyperring case, $\mu_c$ generally increases with $K_2$. However, in the case of $r=3$ and very small $R_{\rm thr}=0.2$, we note that the curve is decreasing at very low $K_2$ which breaks the monotonous structure. This can also be observed in the inset of \cref{fig:k2_v_mu_r=3_Random}(a), where the lower part contains a lighter band, representing slightly higher $\hat R$ values. 
% We attribute this $\hat R$ variability to variations of the hypergraph structure and of $\bm{ \omega}$. 
This slight monotonicity at this extreme desychronization threshold is likely due to a combination of factors including finite-size effects.
% Thus, we claim that this does not invalidate the global conclusion of the existence of a monotonic relation between $\mu_c$ and $K_2$.
Overall, results are robust against variations in the threshold $R_{\rm thr}$.}

\begin{figure}[ht]
    \centering
    \includegraphics[width=0.5\linewidth]{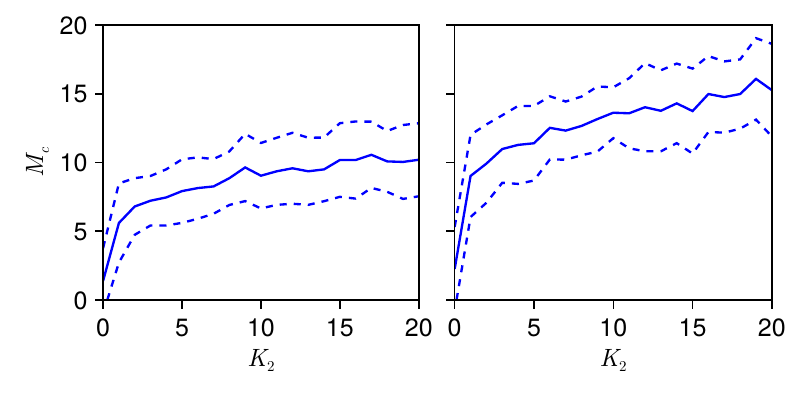}
    \caption{\rev 
    \textbf{The minimal number of controlled nodes $M_c$ required to desynchronize $\hat R$ < 0.4 increases with triadic coupling strength $K_2$ on hyperrings.} 
    We show $M_c$ averaged (solid line) over 50 frequency distributions $\bm{\omega}$, and one standard deviation away (dashed lines).
    We set $N=100$ and $K_1=1$, as in \cref{fig:mu_c}, and $\mu=0.5$. Panel (a): $r=2$; panel (b): $r=3$.}
    \label{fig:M_c}
\end{figure}

\begin{figure}[ht]
    \centering
    \includegraphics[width=0.5\linewidth]{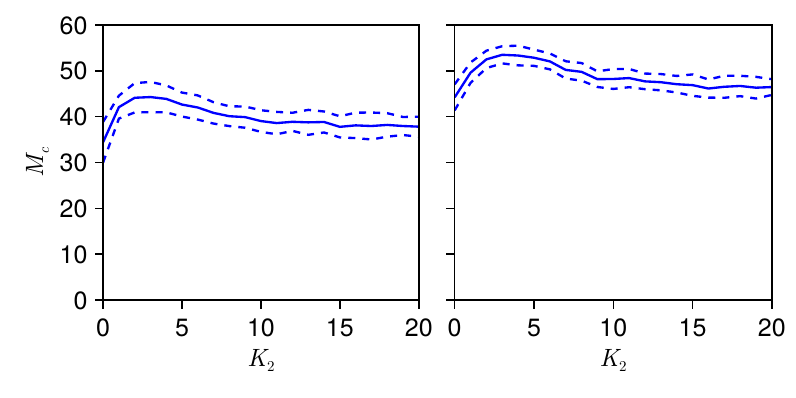}
    \caption{\rev \textbf{The minimal number of controlled nodes $M_c$ required to desynchronize $\hat R$ < 0.4 non-monotically depends on $K_2$ on random hypergraphs.} 
    We show $M_c$ averaged (solid line) over 50 frequency distributions $\bm{\omega}$, and one standard deviation away (dashed lines).
    We set $N=100$, $K_1=1$, $\md1=2r$ and $\md2=2r(r-1)$, as in \cref{fig:mu_c_random}, and $\mu=0.5$. Panel (a): $r=2$; panel (b): $r=3$.}
    \label{fig:M_c_random}
\end{figure}

{\rev
\cref{fig:M_c} and \cref{fig:M_c_random} display a complementary analysis where $\mu=0.5$ is fixed and $M$ varies. They are respectively analogous to \cref{fig:mu_c} and \cref{fig:mu_c_random} and show the minimal required $M_c$ number of controlled nodes so that $\hat R\leq 0.4$.
}

\newpage
\newpage

\subsection{The case of perturbed synchronization as initial conditions}

\begin{figure}[ht]
    \centering
    \includegraphics[width=\linewidth]{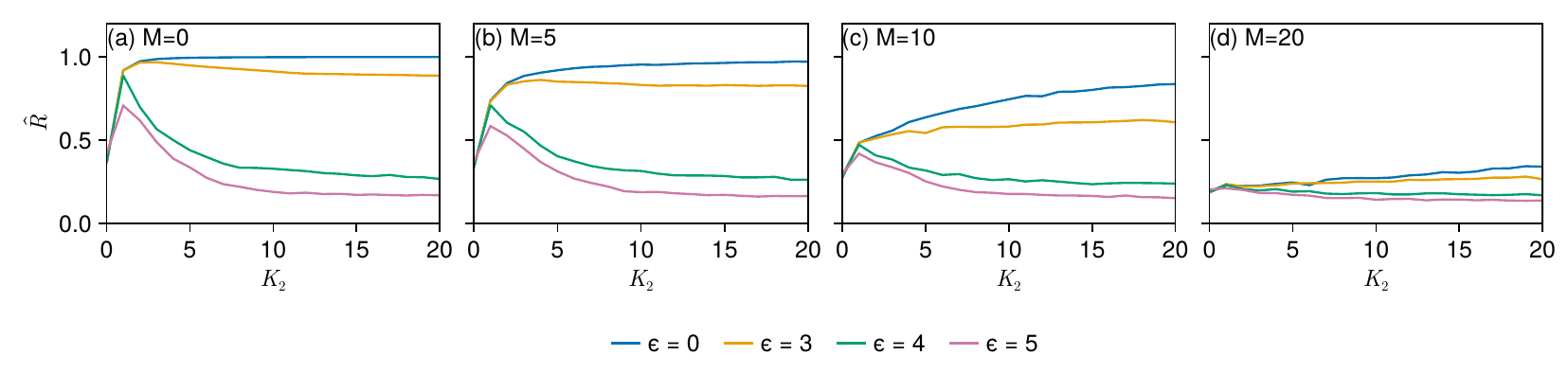}
    %HatR_vs_eps_r_3_N_100_h_0.01_tfinal_200_K1_1_omegaNIter_10.pdf}
    \caption{
    \textbf{Desynchronizing phase oscillators on a hyperring with $r=3$.}
   Similarly to \cref{fig:HatR_r_2_K1_1}: we show $\hat R$ as a function of $K_2$ and $\epsilon$, by controlling (a)-(d) $M=0$, $5$, $10$, and $20$ nodes, respectively.
    Other parameters are set to $N=100$, $r=2$ and $K_1=1$. 
    Order parameter $\hat R$ is averaged over 10 random realizations of frequencies $\bm{\omega}$ and 50 of initial conditions $\theta_0$.
    }
    \label{fig:HatR_r_3_K1_1}
\end{figure}

\begin{figure}[ht]
    \centering
    \includegraphics[width=\linewidth]{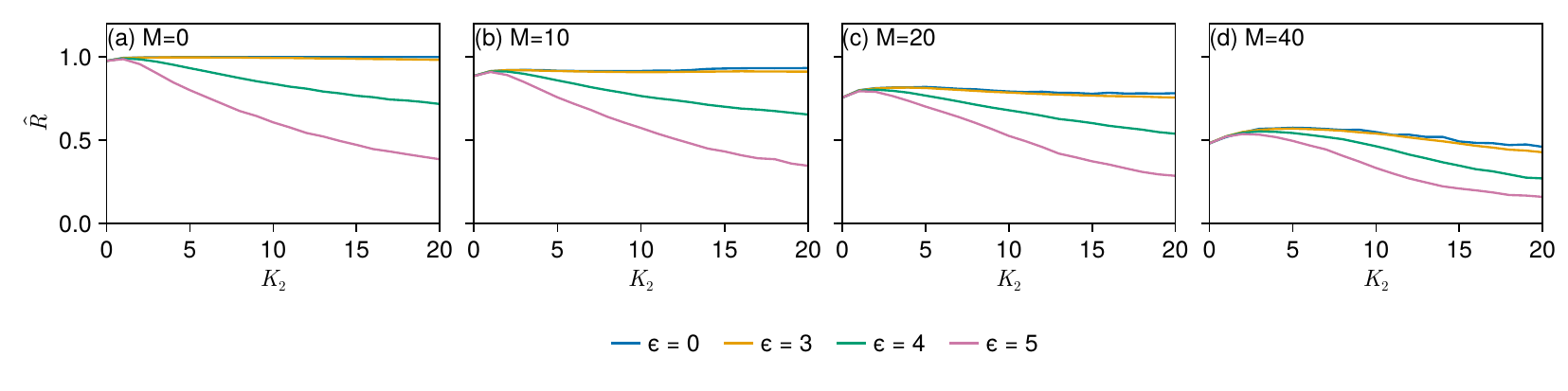}
    \caption{ \textbf{Desynchronizing phase oscillators on Random hypergraphs with $r=3$.}
   Similarly to \cref{fig:HatR_r_2_K1_1}: we show $\hat R$ as a function of $K_2$ and $\epsilon$, by controlling (a)-(d) $M=0$, $5$, $10$, and $20$ nodes, respectively.
    Other parameters are set to $N=100$, $r=2$ and $K_1=1$. 
    Order parameter $\hat R$ is averaged over 50 random realizations of frequencies $\bm{\omega}$, each with a hypergraphs generated via the higher-order Erd\H{o}s-Renyi procedure and 50 of initial conditions $\theta_0$.}
    \label{fig:HatR_Random_r_3_K1_1}
\end{figure}

\cref{fig:HatR_r_2_K1_1,fig:HatR_Random_r_2_K1_1} are respectively analogous to \cref{fig:HatR_r_3_K1_1,fig:HatR_Random_r_3_K1_1} with $r=3$. They display qualitatively equivalent plots.

\end{document}